\documentclass[11pt,a4paper]{article}
\usepackage{diagbox}
\usepackage{jheppub}
\usepackage{graphicx}
\usepackage{amsmath}
\usepackage{amssymb}
\usepackage{amsthm}
\usepackage{mathtools}
\usepackage{appendix}
\usepackage{mathrsfs}
\input epsf
\usepackage{epsfig}
\usepackage{tikz}
\usepackage{subfigure,tikz}
\usepackage{graphics}
\usepackage{graphicx}
\usepackage{diagbox}
\usepackage{mathtools}
\usepackage{xcolor}
\usepackage[thicklines]{cancel}
\usepackage{lscape}
\usepackage{extarrows}

\newcommand{\bea}{\begin{eqnarray}}
	\newcommand{\eea}{\end{eqnarray}}
\newcommand{\bean}{\begin{eqnarray*}}
	\newcommand{\eean}{\end{eqnarray*}}
\newcommand{\nn}{\nonumber \\}

\def\newline{{\hspace{15pt}}}

\def\braket#1{\left\langle #1 \right\rangle}

\def\det{\mathop{\rm det}}

\def\eref#1{(\ref{#1})}
\def\d{{\rm d}}

\def\d{\partial}

\def\what{\widehat}
\def\co{\,,}
\def\ed{\,.}

\newcommand{\parall}[2]{{#1}\ /\kern -0.8em / \  {#2}}

\title{ Tensor Loop Reduction via the Baikov Representation and an Auxiliary Vector }

\author[a]{Liang Zhang}

\affiliation[a]{Zhejiang Institute of Modern Physics, Zhejiang University, Hangzhou, 310027, P. R. China }

\emailAdd{liangzh@zju.edu.cn}

\abstract{In this paper, we introduce a simple and efficient approach for the general reduction of one-loop integrals. Our method employs the introduction of an auxiliary vector and the identification of the tensor structure as an auxiliary propagator. This key insight allows us to express a wide range of one-loop integrals, encompassing both tensor structures and higher poles, in the Baikov representation. By establishing an integral-by-parts (IBP) relation, we derive a recursive formula that systematically solves the one-loop reduction problem, even in the presence of various degenerate cases. Our proposed strategy is characterized by its simplicity and effectiveness, offering a significant advancement in the field of one-loop calculations.  }


\begin{document}
	\maketitle
	\flushbottom
	\section{Introduction} 
	Accurate calculations of higher-loop corrections play a crucial role in achieving precision physics at the Large Hadron Collider (LHC) and future colliders. These calculations are essential for accurately predicting particle interactions and interpreting experimental data. However, the evaluation of loop integrals becomes increasingly challenging as the loop order and complexity of the integrals rise. In particular, the presence of intricate tensor structures and propagators raised to high powers introduces significant computational difficulties.
	
	In recent years, remarkable progress has been made in both the computation and understanding of the analytic structures of scattering amplitudes. Various powerful techniques have emerged to address the reduction of loop integrals at both the integrand and integral levels. Integration-By-Parts (IBP) relations have proven to be highly effective in simplifying loop integrals by relating them to simpler master integrals~\cite{Chetyrkin:1981qh, Tkachov:1981wb, Laporta:2000dsw,vonManteuffel:2012np,vonManteuffel:2014ixa, Maierhofer:2017gsa, Smirnov:2019qkx}. Passarino-Veltman (PV) reduction~\cite{Passarino:1978jh} and Ossola-Papadopoulos-Pittau (OPP) reduction~\cite{Ossola:2006us, Ossola:2007bb, Ellis:2007br} provide alternative approaches to simplify loop integrals, while unitarity-based methods~\cite{Bern:1994zx, Bern:1994cg, Bern:1997sc, Britto:2004nc, Britto:2005ha, Anastasiou:2006gt, Britto:2006sj,Anastasiou:2006jv,Britto:2006fc,Britto:2007tt,Britto:2010um} exploit the cutting equations to derive compact forms of loop integrals. Intersection number techniques~\cite{Mastrolia:2018uzb,Mizera:2019ose,Frellesvig:2019uqt,Frellesvig:2019kgj,Mizera:2019vvs,Frellesvig:2020qot,Caron-Huot:2021xqj} have also emerged as powerful tools for the reduction of loop integrals.
	
	There has been several research~\cite{Feng:2022rwj,Feng:2022iuc,Feng:2022rfz,Li:2022cbx} investigating these general one-loop integrals by contact $\ell$  with an auxiliary vector $R$. Let us see the one-loop  $r$-rank tensor integral with $n$ propagators
	\begin{align}
		I_n^{\mu_1,\ldots,\mu_r}\equiv\int  {d^d \ell \over i(\pi)^{d/2}} {\ell^\mu_1 \ldots \ell^\mu_r \over \prod_{i=1}^{n} D_i } ,
	\end{align}
	where the $i$-th propagator is $D_i=(\ell-q_i)^2-m_i^2$, by introducing an auxiliary vector $R^{\mu}$ 
	\begin{align}
		I^{(r)}_{n}
		\equiv& 2^r I^{\mu_1\cdots \mu_r}_{n} R_{\mu_1}\cdots R_{\mu_r}
		=\int {d^d\ell\over i(\pi)^{d/2}} { (2\ell\cdot R)^r\over D_1\cdots D_n}.
	\end{align}
	The introduction of  $R$ allows for a more concise expression and enhancing the efficiency of reduction in previous studies, such as \cite{Feng:2021enk, Hu:2021nia, Feng:2022uqp, Feng:2022rfz, Chen:2022jux,Feng:2022hyg} for one-loop integrals and \cite{Feng:2022iuc, Chen:2022lue} for higher loops. This technique can be combined with other methods, including  differential operators \cite{Feng:2022rwj,Feng:2022iuc,Feng:2022rfz}, the syzygy equation in Baikov representation \cite{Chen:2022jux, Gluza:2010ws, Larsen:2015ped, Larsen:2016tdk, Zhang:2016kfo, Georgoudis:2016wff, Georgoudis:2017iza, Bohm:2017qme, Bohm:2018bdy, Bendle:2019csk, Boehm:2020zig, Bendle:2021ueg, Schabinger:2011dz}, and  IBP in projective space \cite{Li:2022cbx}. In this paper, we introduce a simple approach that combines  differential operators with respect to  $R$ and the IBP relation  in  Baikov representation. Baikov representation provides a systematic way to express loop integrals in terms of a set of master integrals. After introducing an auxiliary vector and recognizing the tensor structure as a new propagator with negative power, we are able to establish a simple recursive relation for the reduction of general one-loop integrals. This approach proves to be particularly advantageous in handling degenerate cases, where other methods face challenges.
	
	The main objective of this paper is to present our proposed method for the uniform reduction of general one-loop integrals using the Baikov representation and IBP reduction. We demonstrate the effectiveness and simplicity of our approach through several illustrative examples, including the reduction of tadpole, bubble, triangle and pentagon.
	
	The rest of this paper is organized as follows: In Section \ref{review}, we provide a review of two methods for tensor reduction: differential operators and syzygy equations. In Section \ref{ourmethod}, we provide a detailed outline of our method and present the general result for the reduction of one-loop integrals. Section \ref{examples} presents a comprehensive set of examples, showcasing the application of our approach to various types of integrals. Finally, in Section \ref{summary}, we summarize our findings, discuss the implications of our method, and provide an outlook for future research in this area. The paper ends with an appendix. In Appendix \ref{sec:appendix}, we present the results for the pentagon, and compare the computation times of the FIRE6  and our method. There is an ancillary file in which the symbolic results for the pentagon reduction coefficient of rank $r=2$ are given.
	
	\section{Review of two methods}
	\label{review}
	In this paper we mainly consider  the  one-loop $r$-rank tensor integral with $n$ propagators
	\begin{align}
		I^{(r)}_{\boldsymbol{a}_n}
		\equiv\int {d^d \ell\over i (\pi)^{d/2}} { (2\ell\cdot R)^r\over \prod_{j=1}^{n}D_j^{a_j}},
	\end{align}
	where the $j$-th propagator $D_j=(\ell-q_j)^2-m_j^2$, and the loop momentum $\ell$, auxiliary vector $R$, and region momentum $q_j$ exist in $d$-dimensional spacetime, with $q_j=\sum_{i<j}p_i$ and $q_1=0$. In this expression, $\boldsymbol{a}_n=\{a_1,a_2,\ldots,a_n\}$ represents the power list of the $n$ propagators.  Induction of the auxiliary vector can not only simplify the reduction process but also help us to solve the higher-pole case.  One can see any general tensor structure can be recover  by applying differential operators of $R$ on the standard expression. For example,
	\begin{align}
		\int {d^d \ell\over i (\pi)^{d/2}}  {\ell^2 \ell\cdot K\over \prod_{j=1}^n ((\ell-q_j)^2-m_j^2)^{a_j}}\propto (K\cdot \d_R)(\d_R\cdot \d_R)I^{(3)}_{\boldsymbol{a}_n}\ed
	\end{align}
	The more general case of tensor reductions for higher pole can be addressed  by employing differential operators of $m_i^2$,
	\begin{align}\label{diffem}
		\int {d^d \ell\over i (\pi)^{d/2}}  {(2\ell \cdot R)^r\over  \prod_{j=1}^{n}D_j^{a_j}}\propto \left(\prod_i (\d_{m_i^2})^{a_i-1}\right)	\int {d^d \ell\over i (\pi)^{d/2}}  {(2\ell \cdot R)^r\over  \prod_{j=1}^{n}D_j}\ed
	\end{align}
	One can notice that any differential operator of mass can lift the power of the associated propagator by $1$. Given the reduction results for the scalar integral class $I_{\{a_i=2\}}$, where $\{a_i=2\}$ indicates all propagators power $a_j=1$ except $a_i=2$, one can solve the general problem of reducing tensor integrals of higher poles.  Therefore, for the sake of simplicity, we will focus solely on the integrals with simple poles and scalar integrals with single quadratic propagator,  \textit{i.e.}, $I_n^{(r)}\equiv I_{\{1,1,\ldots,1\}}^{(r)}$  and  $I_{\{a_i=2\}}$.
	\subsection{Reduction by differential operators}
	
	In this subsection, we provide a review of utilizing differential operators for tensor reduction. In the original works \cite{Feng:2021enk, Hu:2021nia},   they note that there are two types of differential operators which can lower the rank $r$:
	\begin{align}
		\mathcal{D}_i\equiv q_i\cdot {\d \over \d R},~~i=2,\ldots,n,~~~~ \mathcal{T}\equiv \eta^{\mu\nu}{\d\over \d R^\mu}
		{\d \over \d R^\nu}.\label{def-diffe}
	\end{align}
	It is straightforward to determine the action of these operators
	\begin{align}
		\mathcal{D}_i I^{(r)}_{n}
		& =  r I^{(r-1)}_{n;\widehat{1}}- r I^{(r-1)}_{n;\widehat{i}}+ r  (m^2_1+q^2_i-m^2_i) I^{(r-1)}_{n},  \notag\\
		\mathcal{T} I^{(r)}_{n} 
		& =  4r (r-1)m_1^2 I^{(r-2)}_{n}+ 4r(r-1)I^{(r-2)}_{n;\widehat{1} }. ~~~\label{operator_action}
	\end{align}
	We know that $I^{(r)}_{n}$ can be reduced to master integrals,
	\begin{align}
		I^{(r)}_{n}
		=& \sum_{j=0,\mathbf{b}_j}C^{(r)}_{n\to n;\what{\mathbf{b}_j}}I_{n;\what{\mathbf{b}_j}}\ed \label{reduction-formula}
	\end{align}
	In a straightforward approach, given the results for the integrals $I^{(r')}_{m<n}$, the expressions for $C^{(r)}_{n\to n;\hat{\mathbf{b}}_j}$ can be determined by solving the $n$ partial differential equations in  (\ref{operator_action}), with the property
	\begin{align}
		\mathcal{D}_i
		&=2s_{0i}{\d \over \d s_{00}}+\sum_{j=1}^{n-1} s_{ij}{\d \over \d s_{0j}},   \notag\\
		\mathcal{T}
		&=2D{\d \over \d s_{00}}+4s_{00}{\d^2\over \d s_{00}^2}+4\sum_{i=1}^{n-1}s_{0i}{\d \over \d s_{0i}}{\d \over \d s_{00}}+\sum_{i=1}^{n-1}\sum_{j=1}^{n-1}s_{ij}{\d \over \d s_{0i}}{\d \over \d s_{0j}}\ed ~~~~\label{differential_operators}
	\end{align}
	However, there exists a more efficient approach. The key idea utilized in \cite{Feng:2021enk} is to expand the reduction coefficients based on their tensor structure
	\begin{align}
		C^{(r)}_{n\to n;\what{\mathbf{b}_j}}
		=& \sum_{2a_0+\sum_{k=1}^{n-1}a_k=r}\Bigg\{ c_{n\to n;\what{\mathbf{b}_j}}^{(a_0,\cdots,a_{n-1})}(r) \prod_{k=0}^{n-1} s_{0k}^{a_k}\Bigg\} ,\label{coefficient_expansion_1}
	\end{align}
	where  $s_{00}\equiv R\cdot R, s_{0i}\equiv R\cdot q_{i+1}$.  By substituting this expansion into the $n$ partial differential equations, one can derive $n$ recursion relations for the \textit{expansion coefficients} $c_{n\to n;\hat{\mathbf{b}}j}^{(a_0,\ldots,a_{n-1})}(r)$. Moreover, these recursion relations can be solved through an iterative approach. Finally, by collecting all expansion coefficients, one can compose the desired reduction coefficients. Regarding degenerate cases, these can be addressed through singularity analysis, as shown in \cite{Feng:2022rfz}.
	
	\subsection{Reduction with syzygy in Baikov representation}
	
	The  one-loop integrals in Baikov representation \cite{Baikov:1996iu, Lee:2010wea, Bosma:2017ens, Harley:2017qut, Bosma:2017hrk, Frellesvig:2017aai}, we denote the propagators  and integral as
	\begin{align}
		&z_1=\ell^2-m_1^2\co\ \ \  z_2=(\ell-p_1)^2-m_2^2\co\ \ \  z_3=(\ell-p_1-p_2)^2-m_3^2\co \ldots\co\ \nn
		&z_{n}=(\ell-p_1-\dots-p_E)^2-m_{n}^2\co~~~ z_0=(2\ell \cdot R)\co\nn
		&I^{(r)}_{\boldsymbol{a}_n}\equiv I^{(r)}_{a_1,a_2,\ldots,a_{n}} \equiv   {\mathcal{K}^{-(d-n-1)/2} \over (4\pi)^{n/2}\Gamma((d-n)/2)}\int_{\cal{C}}  d^{n+1}z    \frac{\mathcal{G}(\{z\})^{(d-n-2)/2} z_0^r}{\prod_{i=1}^{n}  z_i^{a_i}}\co
	\end{align}
	where $E=n-1$ is the number of independent external momenta, the Gram determinant $\mathcal{K}$ involving the external momenta do not depend on $\{z\}$ and can be ignored for our subsequent discussions. The  $\mathcal{G}(\{z\})$ is another Gram determinant, which depends on both the loop momentum and the external momenta,
	\begin{align}
		\mathcal{G}(\{z\})=\det G(\ell,p_1,\ldots,p_E,R).
	\end{align}
	Here, the Gram matrix $G$ is defined as
	\begin{align}\label{grammatrix}
		G(q_1,\ldots,q_n) \equiv (q_i \cdot q_j)_{n\times n} \equiv 
		\begin{pmatrix}
			q_1 \cdot q_1 & q_1 \cdot q_2 & \cdots & q_1 \cdot q_n
			\\
			q_2 \cdot q_1 & q_2 \cdot q_2 & \cdots & q_2 \cdot q_n
			\\
			\vdots & & \ddots & \vdots
			\\
			q_n \cdot q_1 & q_n \cdot q_2 & \cdots & q_n \cdot q_n
		\end{pmatrix}\ed
	\end{align}
	Introducing the syzygy module \cite{Chen:2022jux} and considering the IBP relation
	\begin{align}    
		0&=\int d^{n+1}z\sum_{a=0}^{n} \left[\partial_{z_a} \left(P_a{z_0^r \mathcal{G} \{z\}^{(d-n-2)/2}\over \prod_{i=1}^{n} z_i^{b_i}}\right)\right]  \nn
		&=\int d^{n+1}z\sum_{a=0}^{n}\left[ {d-n-2\over 2}P_a\partial_{z_a}\mathcal{G} +(\partial_{z_a}P_a+r z_0^{-1}P_0- b_i z_i^{-1}P_i)  \mathcal{G}\right]{z_0^r \mathcal{G} \{z\}^{((d-2)-n-2)/2}\over \prod_{i=1}^{n} z_i^{b_i}} ,
	\end{align}
	where $P_a$ are polynominals of $z_a$.  It should be noted that the power of $\mathcal{G}$ is dependent on the dimension $d$; therefore, to avoid dimension shifting in the reduction coefficients, it is advantageous to select $P_a$ judiciously to satisfy the syzygy equation.
	\begin{align}
		\sum_{a=0}^{n}\left(P_a\partial_{z_a} \mathcal{G}\right)+P_{n+1}\mathcal{G}=0\ed
	\end{align} 
	Another consideration is that the term $b_i z_i^{-1}P_i$ may increase the power of the $i$-th propagator, which is undesirable. To preclude this, we require $P_i$ to be divisible by $z_i$. With these constraints in place, one can identify solutions to the syzygy equation by searching for syzygy modules $\braket{\partial_{z_0}\mathcal{G},\ldots,\partial_{z_n}\mathcal{G},\mathcal{G}}$ subject to the requirement  $\langle P \rangle \equiv  \langle P_0,P_1,\ldots, P_n \rangle = \langle d_0,d_1,\ldots, d_n \rangle$, where 
	\begin{align}
		d_1=&\{z_1,0,\ldots,0,0,0\}\nn
		\vdots \nn
		d_n=&\{0,0,\ldots,z_n,0,0\}\nn
		d_0=&\{0,0,\ldots,0,1,0\}\nn
		d_{n+1}=&\{0,0,\ldots,0,0,1\}
	\end{align}

	\section{Combined method}
	\label{ourmethod}
	Differential operators and syzygy equation both require extensive algebraic computation. While solving the syzygy equation is typically straightforward at the one-loop level, our objective is to explore a reduction method that does not rely on this approach. Instead, we aim to develop a methodology that can be applied more generally, starting with the one-loop case as a preliminary step. We introduce a simple approach that combines  differential operators with respect to  $R$ and the IBP relation  in  Baikov representation. First, we examine the elements of the Gram matrix given in \eref{grammatrix}. The  Gram matrix contains elements that represent the Lorentz invariant products of loop momenta, external momenta, and auxiliary vectors. Specifically, the matrix elements are constructed from scalar products of the form $\ell\cdot \ell$, $\ell \cdot p_i$, $\ell\cdot R$, $p_i\cdot p_j$, $p_i\cdot R$, and $R\cdot R$, where 
	\begin{align}
		\ell\cdot \ell &= m_1^2+z_1,~~\ell\cdot R = z_0/2, \nn 
		\ell\cdot p_i &= \frac{1}{2}(m_i^2-m_{i+1}^2+z_i-z_{i+1}+p_i \cdot p_i+2\sum_{j=1}^{i-1} p_j\cdot p_i)\ed
	\end{align}
	It is evident that the variables $\{z\}$ only appear in the first column and row of matrix $G$, indicating that $\mathcal{G}$ must be quadratic expressions of $\{z\}$. To simplify the expressions, we introduce the notation
	\begin{align}\label{gformula}
		{\cal G}={\cal G}_{00}z_0^2+\sum_{i=1}^n{\cal G}_{0i}z_0z_i+{\cal G}_{0}z_0+\sum_{i\le j}^n{\cal G}_{ij}z_iz_j+\sum_{i=1}^n{\cal G}_iz_i+{\cal G}_c \ed
	\end{align}
	Since only $\mathcal{G}_{00}, \mathcal{G}_0$ and $\mathcal{G}_c$ are associated with singularities, we will focus exclusively on presenting their specific results.
	\begin{align}\label{gfunc}
		&\mathcal{G}_{00}=-\frac{1}{4}\det \left[{G}(\{z\})_{\bcancel{1(n+1)};\bcancel{1(n+1)}}\right]\co \nn
		&\mathcal{G}_{0}=-\det \left[G(\{z\}=0)_{\bcancel{(n+1)};\bcancel{1}}\right]\co~~
		\mathcal{G}_{c}=\det {G}(\{z\}=0)\co
	\end{align}
	where the $\bcancel{ij};\bcancel{lk}$ indicates  removing the $i$-th and $j$-th rows as well as the $l$-th and $k$-th columns. In addition, the $\{z\}=0$ is to set $z_a=0, \forall a=0,1,\ldots,n$. We consider the integral relation derived from the IBP
	\begin{align}\label{partialz0}
		0&=\int_{\cal{C}} d^{n+1}z ~ \partial_{z_0} \left[   \mathcal{G}(\{z\})^{(d-n-2)/2} \mathcal{G}(\{z\})\frac{z_0^{r-1}}{\prod_{i=1}^{n}  z_i}\right]\nn
		&=\int_{\cal{C}} d^{n+1}z \left[ \Big({d-n\over 2} z_0 \partial_{z_0}\mathcal{G}+(r-1) \mathcal{G}\Big) \times  \mathcal{G}(\{z\})^{(d-n-2)/2} \frac{z_0^{r-2}}{\prod_{i=1}^{n}  z_i}\right]\ed
	\end{align}
	In order to prevent the power of the function $\mathcal{G}$ from changing when differentiating, we append an additional factor of $\mathcal{G}$~\cite{Lee:2014tja}. Since $z_0$ is independent of the propagators, taking the derivative $\partial_{z_0}$ does not increase the power of the propagators. Substituting \eref{gformula} into it and  omitting the summation symbol. Recall that any $z_i$ represents a certain propagator and $z_0$ represents the tensor structure, it is direct to recognize the equation as a reduction relation at integral level  as below 
	\begin{align}\label{basic-relation}
		&(r-1+d-n)\mathcal{G}_{00}I^{(r)}_n+(r-1+{d-n\over 2})\left(\mathcal{G}_{0}I^{(r-1)}_n+\mathcal{G}_{0i}I^{(r-1)}_{n;\widehat{i}}\right)\nonumber\\
		&+(r-1)\left(\mathcal{G}_{c}I^{(r-2)}_n+\mathcal{G}_{i}I^{(r-2)}_{n;\widehat{i}}+\mathcal{G}_{ij}I^{(r-2)}_{n;\widehat{ij}}\right)=0\co
	\end{align}
	where the $\widehat{ij}$ indicates the propagators $z_i$, $z_j$ have been removed. For the case $i\not =j$, the integral $I^{(r)}_{n;\widehat{ij}}$ just represents the sub-sector with two different propagators removed. However, when $i=j$, special care must be taken to keep $z_i$ in the numerator of $I_{n;\widehat{ij}}^{(r-2)}$. The terms on the right-hand side have lower rank than the initial $I_n^{(r)}$. Hence, with the seed scalar integrals, we can use this relation to construct $r$-rank tensor integrals. To enable the recursion relation to progress smoothly, it is critical to properly handle this $i=j$ case. One approach is to transfer the propagator $z_i$ to differential operator directly using the fact $l^2\propto \partial_R\cdot \partial_R,l\cdot q_i\propto q_i\cdot \partial_R$, due to the reduction result of sub-sector is assumed to be known, one can easily apply on any differential operators. Alternatively, we can write $z_i$ as
	\begin{align}\label{dzfunc}
		(z_i-z_{i+1})+z_{i+1}=2\ell\cdot (q_{i+1}-q_i)+f_{i,i+1}+z_{i+1}, 
	\end{align}
	with $n+1\equiv 1, f_{ij}\equiv m_j^2-m_i^2+q_i^2-q_j^2$.
	There is only one linear term in $\ell$, so only one differential operator $\partial_R$ is needed. The $z_{i+1}$ term that appears can be canceled by the denominator. Applying this logic leads to the  simplified result 
	\begin{align}\label{removeii}
		I_{n;\widehat{ii}}^{(r-2)} =& \int\frac{d^d \ell (2\ell \cdot R)^{(r-2)}}{i(\pi)^{d/2}}\frac{z_i}{\prod_{j=1,j\not=i}^{n} z_j} \nn
		=& \int\frac{d^d \ell (2\ell \cdot R)^{(r-2)}}{i(\pi)^{d/2}}\frac{2\ell \cdot (q_{i+1}-q_i)+f_{i,i+1}+z_{i+1}}{\prod_{j=1,j\not=i}^{n} z_j} \nn
		=& {(q_{i+1}-q_i)\cdot \partial_R \over r-1}\int\frac{d^d \ell (2\ell \cdot R)^{(r-1)}}{i(\pi)^{d/2}}\frac{1}{\prod_{j=1,j\not=i}^{n} z_j} \nn 
		&+ \int\frac{d^d \ell (2\ell \cdot R)^{(r-2)}}{i(\pi)^{d/2}}\frac{f_{i,i+1}}{\prod_{j=1,j\not=i}^{n} z_j}+\int\frac{d^d \ell (2\ell \cdot R)^{(r-2)}}{i(\pi)^{d/2}}\frac{z_{i+1}}{\prod_{j=1,j\not=i}^{n} z_j}\nn 
		=& {(q_{i+1}-q_i)\cdot \partial_R \over r-1} I^{(r-1)}_{n;\widehat{i}}+f_{i,i+1} I^{(r-2)}_{n;\widehat{i}}+I^{(r-2)}_{n;\widehat{i,i+1}}\ed
	\end{align}
	Plugging the expression into \eref{basic-relation}, we obtain 
	\begin{align}\label{firstresult}
		\textsf{A}_{n,r} I_n^{(r)} +\textsf{B}_{n,r} I_n^{(r-1)} +\textsf{C}_{n,r} I_n^{(r-2)} +\textsf{B}_{n,r;\widehat{i}} I_{n;\widehat{i}}^{(r-1)}+\textsf{C}_{n,r;\widehat{i}}I_{n;\widehat{i}}^{(r-2)}+\textsf{C}_{n,r;\widehat{ij}}I_{n;\widehat{ij}}^{(r-2)}=0\co
	\end{align}
	where the  coefficients are
	\begin{align}
		&\textsf{A}_{n,r}=(r-1+d-n)\mathcal{G}_{00}\co\\
		&\textsf{B}_{n,r}=(r-1+{d-n\over 2}){\cal G}_{0},~ \textsf{B}_{n,r;\widehat{i}}=(r-1+{d-n\over 2}){\cal G}_{0i}+{\cal G}_{ii}{(q_{i+1}-q_i)\cdot \partial_R}\co\\
		&\textsf{C}_{n,r}=(r-1){\cal G}_{c},~ \textsf{C}_{n,r;\widehat{i}}=(r-1)({\cal G}_{i}+f_{i,i+1} {\cal G}_{ij}),~ \textsf{C}_{n,r;\widehat{ij}}=(r-1) ({\cal G}_{ij}+\delta_{j,i+1}{\cal G}_{ii})\ed \label{cfactor}
	\end{align}
	
	In fact, all degenerate cases are captured by $ \mathcal{G}_{00},{\cal G}_{0}, {\cal G}_{c}$, as we discussed below.
	For the non-degenerate case $\mathcal{G}_{00}\not =0$, dividing the \eref{firstresult} by the prefactor $\textsf{A}_{n,r}$, and introducing simplified notations, we can derive the recursion relation that governs $I_n^{(r)}$
	\begin{align}\label{secondresult}
		I_n^{(r)} ={-1\over \textsf{A}_{n,r}}\left(\textsf{B}_{n,r}^{-} I_n^{(r-1)} +\textsf{C}_{n,r} I_n^{(r-2)} +\textsf{B}_{n,r;\widehat{i}} I_{n;\widehat{i}}^{(r-1)}+\textsf{C}_{n,r;\widehat{i}}I_{n;\widehat{i}}^{(r-2)}+\sum_{i<j}\textsf{C}_{n,r;\widehat{ij}}I_{n;\widehat{ij}}^{(r-2)}\right) \ed
	\end{align}
	
	As we can see, when $\mathcal{G}_{00}=0$, i,e. $\textsf{A}_{n,r}$, the \eref{secondresult} is no longer applicable. We can divide both sides of \eref{firstresult} by $\textsf{B}_{n,r}$ and shift $r-1$ to $r$, resulting in the following expression
	\begin{align}\label{thirdresult}
		I_n^{(r)} = {-1\over \textsf{B}_{n,r+1}}\left(\textsf{C}_{n,r+1} I_n^{(r-1)} +\textsf{B}_{n,r+1;\widehat{i}}I_{n;\widehat{i}}^{(r)}+\textsf{C}_{n,r+1;\widehat{i}}I_{n;\widehat{i}}^{(r-1)}+\sum_{i<j}\textsf{C}_{n,r+1;\widehat{ij}}I_{n;\widehat{ij}}^{(r-1)}\right)\ed
	\end{align}
	Setting $r=0$ leads all $\textsf{C}_{n}$ to vanish, as evident from (\ref{cfactor}). This implies that $I_n$ is no longer a master integral.
	
	When the conditions $\mathcal{G}_{00}=0$ and $\mathcal{G}_{0}=0$ are satisfied, meaning $\textsf{A}_{n,r}=0$ and $\textsf{B}_{n,r}=0$, the previous recursion relation is no longer valid. To address this breakdown, we can divide both sides of \eref{firstresult} by the prefactor $\textsf{C}_{n,r}$, and perform a shift of replacing $r-2$ by $r$. This gives the result
	\begin{align}\label{forthresult}
		I_n^{(r)} = {-1\over \textsf{C}_{n,r+2}}\left(\textsf{B}_{n,r+2;\widehat{i}} I_{n;\widehat{i}}^{(r+1)}+\textsf{C}_{n,r+2;\widehat{i}}I_{n;\widehat{i}}^{(r)}+\sum_{i<j}\textsf{C}_{n,r+2;\widehat{ij}}I_{n;\widehat{ij}}^{(r)}\right)\ed
	\end{align}
	
	Then we turn to discuss the reduction for higher pole case $I_{\{a_i=2\}}$. In order to lift the power of propagators,  one can consider using $\partial_{z_i}$,  then we translate  $z_i$'s  appearing in the numerator to the differential operators of $R$  acting on the standard  tensor integrals $I_n^{(r)}$ we have obtained.  A little difference here is that there is no need to introduce $z_0$  anymore.  The scalar  one-loop integral in Baikov representation without tensor structure is 
	\begin{align}
		I_{\{1,\ldots,1\}} \equiv    C_n(d) \mathcal{K}^{-(d-n)/2} \int_{\cal{C}} d^{n}z ~ \mathcal{G}^{scalar}(\{z\})^{(d-n-1)/2}    \frac{1 }{\prod_{i=1}^{n}  z_i }\co
	\end{align}
	where the Gram determinant is $\mathcal{G}^{scalar}(\{z\})=\det G(\ell,p_1,\ldots,p_E)$. Analogous to \eref{partialz0}, 	 
	\begin{align}
		0&=\int_{\cal{C}} d^{n}z ~ \partial_{z_i} \left[   \mathcal{G}^{scalar}(\{z\})^{(d-n-1)/2} \mathcal{G}^{scalar}(\{z\})\frac{1}{\prod_{i=1}^{n}  z_i}\right]\nn
		&=\int_{\cal{C}} d^{n}z \left[ \Big({d-n+1\over 2} \partial_{z_i}\mathcal{G}^{scalar}-{ \mathcal{G}^{scalar}\over z_i}\Big) \times  \mathcal{G}^{scalar}(\{z\})^{(d-n-1)/2} \frac{1}{\prod_{i=1}^{n}  z_i}\right]\ed \label{standone}
	\end{align}
	Based on the definition of  $G(\ell,p_1,\ldots,p_E)$, we know $\mathcal{G}^{scalar}(\{z\})$ is a quadratic expression of $\{z\}$,
	\begin{align}
		{\cal G}^{scalar}= \sum_{j\le k}^n{\cal G}_{jk}^{scalar}z_jz_k+\sum_{j}\mathcal{G}_j^{scalar} z_j+{\cal G}^{scalar}_c \ed
	\end{align}
	Plugging it into \eref{standone},
	\begin{align}\label{highrecur}
		&\textsf{H}^{i+}_n I_{\{a_i=2\}} + \textsf{H}_n I_{n}  + \textsf{H}_{n;\widehat{i}} I_{\{a_i=0\}}\nn
		&+\sum_{j\neq i}\left(\textsf{H}_{n;\widehat{j}} I_{\{a_j=0\}}+ \textsf{H}^{i+}_{n;\widehat{jj}} I_{\{a_i=2,a_j=-1\}}+\textsf{H}^{i+}_{n;\widehat{j}} I_{\{a_i=2,a_j=0\}}\right)=0 \ed
	\end{align}
	Paralleling the approach in \eref{removeii},
	\begin{align}
		I_{\{a_i=2,a_j=-1\}}=f_{ji}I_{\{a_i=2,a_j=0\}}+I_{\{a_j=0\}}+(q_i-q_j)\cdot \d_R I_{\{a_i=2,a_j=0\}}^{(1)} \ed
	\end{align}
	Plugging it into \eref{highrecur}
	\begin{align}\label{highrecur2}
		&\textsf{H}^{i+}_n I_{\{a_i=2\}} + \textsf{H}_n I_{n}  + \textsf{H}_{n;\widehat{i}} I_{\{a_i=0\}} +\sum_{j\neq i}\Big[\left(\textsf{H}_{n;\widehat{j}}+\textsf{H}^{i+}_{n;\widehat{jj}}\right) I_{\{a_j=0\}}\nn
		&+ \Big(f_{ji}\textsf{H}^{i+}_{n;\widehat{jj}}+\textsf{H}^{i+}_{n;\widehat{j}}\Big) I_{\{a_i=2,a_j=0\}}+\textsf{H}^{i+}_{n;\widehat{jj}}(q_i-q_j)\cdot \d_R I_{\{a_i=2,a_j=0\}}^{(1)}\Big]=0 ,
	\end{align}
	where the coefficients are
	\begin{align}
		\textsf{H}^{i+}_n&=-\mathcal{G}^{scalar}_c,~ \textsf{H}_n={d-n-1\over 2}\mathcal{G}^{scalar}_i,~     \textsf{H}_{n;\widehat{i}}=(d-n)\mathcal{G}^{scalar}_{ii}, \nn
		\textsf{H}^{i+}_{n;\widehat{j}}&=-\mathcal{G}^{scalar}_{j},~ \textsf{H}_{n;\widehat{j}}={d-n-1\over 2}\mathcal{G}^{scalar}_{ij},~ \textsf{H}^{i+}_{n;\widehat{jj}}=-\mathcal{G}^{scalar}_{jj}.
	\end{align}
The recursion relation displayed in \eqref{highrecur2} demonstrates that higher-pole integrals $I_{{a_i=2}}$ can be expressed in terms of the master integral $I_n, I_{{a_i=0}}$, and lower-sector integrals $I_{{a_i=2,a_j=0}}$, whose values are known from previous recursion relations.

	\section{Examples}
	\label{examples}
	To enhance clarity and conciseness in presenting the reduction coefficients, we will adopt specific symbols throughout this paper,
	\begin{align}
		\quad s_{00}\equiv R \cdot R, \quad s_{0i}\equiv R \cdot q_{i+1}, \quad s_{ij}\equiv q_{i+1}\cdot q_{j+1}, \quad \forall i,j =1,\ldots,n.
	\end{align}
	The upcoming Tadpoles and Bubbles subsections will present step-by-step calculations exemplifying our approach. By walking through specific examples, we aim to demonstrate the utilization of the method in practice for computing integral families of interest.
	\subsection{Tadpoles}
	In this subsection, we examine the reduction of tensor tadpoles, considering the simplest case. To begin, let's explicitly define the propagators involved,
	\begin{align}
		z_1 = \ell^2 - m_1^2\co \quad z_0 = 2\ell\cdot R\ed
	\end{align}
	We can express the polynomial $\mathcal{G}$ in the Baikov presentation as follows,
	\begin{align}
		\mathcal{G} = \det\begin{pmatrix}
			m_1^2+z_1  & z_0/2\\
			z_0/2 & s_{00}
		\end{pmatrix} = -\frac{z_0^2}{4}+s_{00}z_1+s_{00}m_1^2\ed
	\end{align}
	Using this expression, we can compute the derivative of $\mathcal{G}$ with respect to $z_0$
	\begin{align}
		\frac{\partial \mathcal{G}}{\partial z_0} = -\frac{z_0}{2}\ed
	\end{align}
	Consequently, we can rewrite the integrand of \eref{partialz0} as
	\begin{align}
		\left(\textsf{A}_{1,r}\frac{z_0^r}{z_1} + \textsf{C}_{1,r} \frac{z_0^{r-2}} {z_1} + \textsf{C}_{1,r;\widehat{1}} z_0^{r-2}\right)\mathcal{G}^{(d-3)/2} = 0\ed
	\end{align}
	In the above equation, we have the coefficients,
	\begin{align}
		\textsf{A}_{1,r}=-\frac{d+r-2}{4}, \quad \textsf{C}_{1,r}=(r-1)m_1^2 s_{00}\ed
	\end{align}
	The last term in the equation is zero, as there are no propagators in it. As a result, we obtain the following relation,
	\begin{align}
		I_1^{(r)}=\frac{4(r-1)m_1^2s_{00}}{d+r-2}I_1^{(r-2)}\ed
	\end{align}
	Consequently, the final result for $I_1^{(r)}$ can be summarized as,
	\begin{align}\label{bub}
		I_1^{(r)}=\left\{ 
		\begin{aligned}
			&\qquad  0 ~~~~~~~~~~~~~~~~~~~~~~~~~~~~~ r=odd,  \\
			&\frac{2^r(r-1)!!m_1^r R^r}{(d+r-2)!!/(d-2)!!}I_1 ~~~~~r=even.
		\end{aligned}
		\right.
	\end{align}
	The results for non-standard tadpole integrals can be obtained through  momentum shifting
	\begin{align}
		\int d^d\ell {(2 \ell\cdot R)^r \over (\ell-K)^2-m^2}&\xlongequal{\ell \to \ell + K}\int d^d\ell {((2 \ell+2 K)\cdot R)^r \over \ell^2-m^2}\nn
		&=\sum_{a=0}^{r}{r \choose a}(2R\cdot K)^a\int d^d\ell {(2 \ell\cdot R)^{r-a} \over \ell^2-m^2}.
	\end{align}
	In order to elucidate the methodology, we consider $r=1$ as an example
	\begin{align}
		\int d^d \ell {(2 \ell\cdot R) \over (\ell-K)^2-m^2} = I_1^{(1)}+(2R\cdot K)I_1=(2R\cdot K)I_1 .
	\end{align}
	\subsection{Bubbles}
	
	Bubbles is the simplest case involving singularities. The explicit propagators are given by
	\begin{align}
		z_1=\ell^2-m_1^2, \quad z_2=(\ell-q_2)^2-m_2^2, \quad z_0=2\ell\cdot R\ed
	\end{align}
	The polynomial $\mathcal{G}$ in the Baikov presentation is defined as,
	\begin{align}
		\mathcal{G} = \det\begin{pmatrix}
			m_1^2+z_1 & -(m_1^2-m_2^2+s_{11}+z_1-z_2)/2 & z_0/2 \\
			-(m_1^2-m_2^2+s_{11}+z_1-z_2)/2 & s_{11} & s_{01} \\
			z_0/2 & s_{01} & s_{00}
		\end{pmatrix}\ed
	\end{align}
	The derivative of $\mathcal{G}$ with respect to $z_0$ is given by
	\begin{align}
		\partial_{z_0} \mathcal{G}=-s_{01}(m_1^2-m_2^2+s_{11}+z_1-z_2)-2s_{11} z_0\ed
	\end{align}
	Substituting this expression into  \eref{partialz0}, we obtain:
	\begin{align} \label{buble-basic}
		\textsf{A}_{2,r} I_n^{(r)} +\textsf{B}_{2,r} I_2^{(r-1)} +\textsf{C}_{2,r} I_2^{(r-2)} + \textsf{B}_{2;\widehat{i}}^{(r-1)} I_{2;\widehat{i}}^{(r-1)}+\textsf{C}_{2,r;\widehat{i}} I_{2;\widehat{i}}^{(r-2)}+\textsf{C}_{2,r;\widehat{ij}} I_{2;\widehat{ij}}^{(r-2)}=0 \ed
	\end{align}
	In order to derive expressions for $I_{2;\widehat{1,1}}^{(r-2)}$ and $I_{2;\widehat{2,2}}^{(r-2)}$, we utilize \eref{dzfunc} to write $z_1$ and $z_2$ in the following form
	\begin{align}
		z_1 = m_2^2 - m_1^2 - s_{11} + 2\ell \cdot q_2 + z_2 = z_2 + m_2^2 - m_1^2 - s_{11} + q_2 \cdot \partial_R z_0\co \\
		z_2 = m_1^2 - m_2^2 + s_{11} - 2\ell \cdot q_2 + z_1 = z_1 + m_1^2 - m_2^2 + s_{11} - q_2 \cdot \partial_R z_0\ed
	\end{align}
	Substituting these equations, we can express the terms $I_{2;\widehat{11}}^{(r-2)}$ and $I_{2;\widehat{22}}^{(r-2)}$ as follows,
	\begin{align}
		I_{2;\widehat{11}}^{(r-2)} &= (m_2^2 - m_1^2 - s_{11})I_{2;\widehat{1}}^{(r-2)} + \frac{q_2 \cdot \partial_R}{(r-1)}I_{2;\widehat{1}}^{(r-1)} + I_{2;\widehat{12}}^{(r-2)}\co \\
		I_{2;\widehat{22}}^{(r-2)} &= -(m_2^2 - m_1^2 - s_{11})I_{2;\widehat{2}}^{(r-2)} - \frac{q_2 \cdot \partial_R}{(r-1)}I_{2;\widehat{2}}^{(r-1)} + I_{2;\widehat{12}}^{(r-2)}\ed
	\end{align}
	Then, we can get the bubble version of \eref{secondresult}
	\begin{align}\label{nopole}
		I_2^{(r)} ={-1\over A_{2,r}}\left(A_{2,r}^{-} I_2^{(r-1)} +A_{2,r}^{--} I_2^{(r-2)} +\widetilde{A}_{2,r;\widehat{i}}^{-} I_{2;\widehat{i}}^{(r-1)}+\widetilde{A}_{n,r;\widehat{i}}^{--}I_{2;\widehat{i}}^{(r-2)}\right)\ed
	\end{align}
	Next, we provide a detailed expression for  \eref{gfunc}
	\begin{align}
		&\mathcal{G}(\{z\})_{\bcancel{13};\bcancel{13}} = \det\begin{pmatrix}
			s_{11}
		\end{pmatrix}, \quad
		\mathcal{G}(\{z\}=0)_{\bcancel{3};\bcancel{1}} = \det\begin{pmatrix}
			- \frac{1}{2}(m_1^2+m_2^2-s_{11}) & 0 \\
			s_{11} & s_{01}
		\end{pmatrix}\co \nonumber\\
		&\mathcal{G}(\{z\}=0)= \det\begin{pmatrix}
			m_1^2 & -(m_1^2-m_2^2+s_{11})/2 & 0 \\
			-(m_1^2-m_2^2+s_{11})/2 & s_{11} & s_{01} \\
			0 & s_{01} & s_{00}
		\end{pmatrix}\ed
	\end{align}
	The specific components of $\mathcal{G}$ are given by
	\begin{align}
		&\mathcal{G}_{00}= -\frac{1}{4}s_{11}\co\\
		&\mathcal{G}_{0}= \frac{1}{2} \left(m_1^2-m_2^2+s_{11}\right) s_{01}\co \\
		&\mathcal{G}_{c}=-\frac{1}{4} \left((m_1^2 - m_2^2)^2s_{00}-(2m_1^2+2m_2^2-s_{11})s_{11}s_{00}+4m_1^2s_{01}^2 \right)\co\\
		& \mathcal{G}_{01}=\frac{1}{2} s_{01}\co \quad \mathcal{G}_{02}=-\frac{1}{2} s_{01}\co \\
		& \mathcal{G}_{1}=-\frac{1}{2}  \left(\left(m_1^2 -m_2^2 -s_{11} \right)s_{00}+2 s_{01}^2\right)\co \quad \mathcal{G}_{2}=\frac{1}{2}  \left(m_1^2-m_2^2+s_{11}\right) s_{00}\co \\
		&\mathcal{G}_{11}= -\frac{1}{4} s_{00}, \quad  \mathcal{G}_{12}= \frac{1}{2} s_{00}, \quad \mathcal{G}_{22}= -\frac{1}{4} s_{00}\ed
	\end{align}
	Then, we give the explicitly expression of \eref{nopole}
	\begin{align}\label{nopoleexp}
		I_2^{(r)}=& { (d+2r-4)(m_1^2-m_2^2+s_{11})s_{01}  \over (d+r-3)s_{11}}     I_2^{(r-1)} \nn
		&+{r-1 \over (d+r-3)s_{11}}\Big[\left((2m_1^2+2m_2^2-s_{11})s_{11}s_{00}-(m_1^2 - m_2^2)^2s_{00}-4m_1^2s_{01}^2\right)  I_2^{(r-2)} \nn
		&+(m_1^2-m_2^2+s_{11}) s_{00}  I_{2;\widehat{2}}^{(r-2)}   + ((m_2^2-m_1^2+3s_{11})s_{00}-4s_{01}^2) I_{2;\widehat{1}}^{(r-2)}\Big]\nn
		&+{(d+2r-4)s_{01}+s_{00}q_2\cdot\partial_R \over (d+r-3)s_{11}}\left( I_{2;\widehat{1}}^{(r-1)}- I_{2;\widehat{2}}^{(r-1)}\right).
	\end{align}
	Clearly, we can reduce $I_2^{(r)}$ to $I_2$ and $I_{2;\widehat{i}}$ through repeatedly applying \eref{nopoleexp}
	\begin{align}\label{bubhighrank}
		I_2^{(r)} = C_{2}^{(r)}I_2 + C_{2;\widehat{2}}^{(r)}I_{2;\widehat{2}} + C_{2;\widehat{1}}^{(r)}I_{2;\widehat{1}}\ed
	\end{align}
	Here are the expressions for $C_2$, $C_{2;\widehat{2}}$, and $C_{2;\widehat{1}}$ for different values of $r$:
	\begin{itemize}
		\item $r=1$
		\begin{align}\label{eq:bubble-rank1}
			& C_{2}^{(1)}={s_{01} (m_1^2-m_2^2+s_{11}) \over{s_{11}}}\co \\
			& C_{2;\widehat{1}}^{(1)}=\frac{s_{01}}{s_{11}}\co \quad C_{2;\widehat{2}}^{(1)}= - \frac{s_{01}}{s_{11}}\ed \label{requal1}
		\end{align}
		\item $r=2$
		\begin{align}
			C_{2}^{(2)}&=\frac{\left(\left(m_1^2-m_2^2\right){}^2-s_{11} \left(2 m_1^2+2 m_2^2-s_{11}\right)\right) \left(s_{11}  s_{00}-d s_{01}^2\right)}{(1-d) s_{11}^2}+\frac{4  m_1^2  s_{01}^2}{ s_{11}}\co\\
			C_{2;\widehat{2}}^{(2)}&=\frac{\left(m_1^2-m_2^2+s_{11}\right) \left(s_{11} s_{00}-d s_{01}^2\right)}{(d-1) s_{11}^2}\co\\
			C_{2;\widehat{1}}^{(2)}&=\frac{\left(-m_1^2+m_2^2+s_{11}\right) \left(s_{11} s_{00}-d s_{01}^2\right)+4 (d-1) s_{11} s_{01}^2}{(d-1) s_{11}^2}\ed
		\end{align}
		\item $r=3$
		\begin{align}
			C_{2}^{(3)}=&\frac{\left(-\left(m_2^2-s_{11}\right){}^2-m_1^4+2 m_2^2 m_1^2\right) \left(m_1^2-m_2^2+s_{11}\right) s_{01} \left(3 s_{11} s_{00}-(d+2) s_{01}^2\right)}{(d-1) s_{11}^3}\nonumber\\
			&+\frac{2 m_1^2 \left(m_1^2-m_2^2+s_{11}\right) s_{01} \left((d-4) s_{01}^2+3 s_{11} s_{00}\right)}{(d-1) s_{11}^2}\co\\
			C_{2;\widehat{2}}^{(3)}=&\frac{\left(\left(m_2^2-s_{11}\right){}^2+m_1^4-2 m_2^2 m_1^2\right) s_{01} \left(3 s_{11} s_{00}-(d+2) s_{01}^2\right)}{(d-1) s_{11}^3}\nonumber\\
			&-\frac{2 m_1^2 s_{01} \left(\left(d^2-2 d+4\right) s_{01}^2+3 (d-2) s_{11} s_{00}\right)}{(d-1) d s_{11}^2}\co\\
			C_{2;\widehat{1}}^{(3)}=&\frac{s_{01} \left(-4 (d-1) m_2^2 s_{11}+d m_1^4-2 d m_2^2 m_1^2+d m_2^4\right) \left((d+2) s_{01}^2-3 s_{11} s_{00}\right)}{(d-1) d s_{11}^3}\nonumber\\
			&+\frac{s_{01} \left(4 (d-1) d m_1^2 s_{11} s_{01}^2+d s_{11}^2 \left((7 d-10) s_{01}^2+3 s_{11} s_{00}\right)\right)}{(d-1) d s_{11}^3}\ed
		\end{align}
	\end{itemize}
	
	\subsubsection{$\mathcal{G}_{00}=0$}
	
	In this subsection, it is important to note that the $I_2^{(r)}$ integrals discussed here are modified versions that have been adjusted to the same limit. As we can see, when $\mathcal{G}_{00}=0$, i,e. $\textsf{A}_{2,r}=0$, the \eref{nopole} is no longer applicable. We can divide both sides of Equation (\ref{buble-basic}) by $\textsf{B}_{2,r}=0$ and shift $r-1$ to $r$, resulting in the following expression
	\begin{align}
		I_2^{(r)} =&{-1\over \textsf{B}_{2,r+1}}\left(\textsf{C}_{2,r+1} I_2^{(r-1)} +\textsf{B}_{2,r+1;\widehat{i}}I_{2;\widehat{i}}^{(r)}+\textsf{C}_{n,r+1;\widehat{i}}I_{2;\widehat{i}}^{(r-1)}\right)\nn
		= &{ r \over  (d+2r-2)(m_1^2-m_2^2)s_{01}}\Big[\left((m_1^2 - m_2^2)^2s_{00}+4m_1^2s_{01}^2 \right)  I_2^{(r-1)} \nn
		&-(m_1^2-m_2^2)  s_{00} I_{2;\widehat{2}}^{(r-1)}   - \left((m_2^2-m_1^2)s_{00}-4s_{01}^2\right) I_{2;\widehat{1}}^{(r-1)}\Big]\nn
		&-{(d+2r-2)s_{01}+s_{00}q_2\cdot\partial_R \over (d+2r-2)(m_1^2-m_2^2)s_{01}}\left( I_{2;\widehat{1}}^{(r)}- I_{2;\widehat{2}}^{(r)}\right) \ed
	\end{align}
	It is evident that $r=0$ can be computed, which implies that $I_2$ can be represented in terms of $I_{2;\widehat{2}}$ and $I_{2;\widehat{1}}$. Consequently, $I_2$ does not appear in the final result, and we have
	\begin{align}\label{q2limit}
		I_2^{(r)} = C_{2;\widehat{2}}^{(r)}I_{2;\widehat{2}}+C_{2;\widehat{1}}^{(r)}I_{2;\widehat{1}}\ed
	\end{align}
	Next, we provide the specific expressions for $C_{2;\widehat{1}}$ and $C_{2;\widehat{2}}$ for different values of $r$:
	\begin{itemize}
		\item $r=0$
		\begin{align}
			C_{2;\widehat{1}}=-\frac{1}{m_1^2-m_2^2} \co \quad C_{2;\widehat{2}}=\frac{1}{m_1^2-m_2^2}\co
		\end{align}
		\item $r=1$
		\begin{align}
			C_{2;\widehat{1}}^{(1)}=\frac{2 s_{01} \left((d-2) m_2^2-d m_1^2\right)} {d \left(m_1^2-m_2^2\right)^2}\co\quad C_{2;\widehat{2}}^{(1)}=\frac{4 s_{01} m_1^2} {d\left(m_1^2-m_2^2\right)^2}\co
		\end{align}
		\item $r=2$
		\begin{align}\label{factor}
			C_{2;\widehat{1}}^{(2)}&=-\frac{4 \left(-2 \left(d^2-4\right) m_2^2 m_1^2+d (d+2) m_1^4+(d-2) d m_2^4\right) s_{01}^2}{d (d+2) \left(m_1^2-m_2^2\right){}^3}-\frac{4 m_2^2 s_{00}}{d m_1^2-d m_2^2}\co\nonumber\\
			C_{2;\widehat{2}}^{(2)}&=\frac{32 m_1^4 s_{01}^2}{d (d+2) \left(m_1^2-m_2^2\right){}^3}+\frac{4 m_1^2 s_{00}}{d m_1^2-d m_2^2}\ed
		\end{align}
	\end{itemize}
	
	\subsubsection{$\mathcal{G}_{00}=0$ and $\mathcal{G}_{0}=0$}
	
	There is still another pole $m_1=m_2=m$. In this case, $\textsf{B}_{2,r}=0$. By dividing both sides of \eref{buble-basic} by $\textsf{C}_{2,r}=0$ and shifting $r-2$ to $r$, we obtain the following expression
	\begin{align}
		I_2^{(r)} =&{-1\over \textsf{C}_{2,r+2}}\left(\textsf{B}_{2,r+2;\widehat{i}}I_{2;\widehat{i}}^{(r+1)}+\textsf{C}_{n,r+2;\widehat{i}}I_{2;\widehat{i}}^{(r)}\right)\nn
		=& { d s_{01}+s_{00}q_2\cdot\partial_R \over 4m^2s_{01}^2 }\left( I_{2;\widehat{1}}^{(r+1)}- I_{2;\widehat{2}}^{(r+1)}\right)-{1\over m^2 }I_{2;\widehat{1}}^{(r)} \ed
	\end{align}
	In this scenario, $I_{2;\widehat{1}}$ is equal to $I_{2;\widehat{2}}$. Therefore, the final result can be expressed as
	\begin{align}
		I_2^{(r)}= C_{2;\widehat{1}}^{(r)}I_{2;\widehat{1}}\ed
	\end{align}
	Now, we provide the specific expressions for $C_{2;\widehat{1}}$ for different values of $r$:
	\begin{itemize}
		\item $r=0$
		\begin{align}\label{highpole}
			C_{2;\widehat{1}}=\frac{d-2}{2 m^2} \co
		\end{align}
		\item $r=1$
		\begin{align}
			C_{2;\widehat{1}}^{(1)}=\frac{(d-2)s_{01}}{2m^2}\co
		\end{align}
		\item $r=2$
		\begin{align}
			C_{2;\widehat{1}}^{(2)} = \frac{2}{3} \left(\frac{(d-2) s_{01}^2}{m^2}+3 s_{00}\right)\ed
		\end{align}
	\end{itemize}
	
	Interestingly, high-pole tadpole integrals can be found from $r=0$ integral. Specifically, when the momentum $q_2\to 0$ and mass $m_2\to m_1$, the second bubble propagator $z_2$ degenerates into $z_1$. In other words, the bubble integral transforms into a higher-pole tadpole integral, i.e., $I_{\{1,1\}} \to I_{\{2\}}$. Applying \eref{highpole} with $q_2=0$ then yields a recurrence relation directly connecting the higher pole bubble integral to the tadpole base case
	\begin{align}\label{n=0}
		I_{\{2\}}=\frac{d-2}{2m_1^2}I_{\{1\}}\ed
	\end{align}
	For $r$ is odd, $I_{\{a_1\}}^{(r)}=0$ which can be followed from  \eref{bub}. Specially,
	\begin{align}
		I_{\{2\}}^{(1)}=0\ed
	\end{align}
	For $r$ is even, 
	\begin{align}
		I_{\{1\}}^{(r)}=\frac{2^r(r-1)!!m_1^r R^r}{(d+r-2)!!/(d-2)!!}I_{\{1\}}.
	\end{align}
	Arbitrary powers of the tadpole propagator $z_1$ can be generated by repeatedly applying the differential operator $\partial_{m_1^2}$. For example,
	\begin{align}
		I_{\{2\}}^{(r)}=\frac{2^{r-1}(r-1)!!m_1^{r-2} R^r}{(d+r-2)!!/(d-2)!!}\left( r I_{\{1\}} + 2 m_1^2 I_{\{2\}}\right).
	\end{align}
	Substituting  \eref{n=0} into it, then
	\begin{align}
		I_{\{2\}}^{(r)}=\frac{2^{r-1}(r-1)!!m_1^{r-2} R^r}{(d+r-2)!!/(d-2)!!}( r + d-2 )I_{\{1\}}\ed
	\end{align}
	
	\subsubsection{${\rm Det} ~Q=0$}
	
	 As stated in \cite{Li:2022cbx}, $\det Q=0$ is also a degenerate case where $Q_{ij}=(m_i^2+m_j^2-(q_i-q_j^2))/2$). But it cannot be well handled in our algorithm, we can use their scalar reduction result as an input to get the tensor reduction. In bubble integral $\det Q=0$ gives $s_{11}=(m_2\pm m_1)^2$. Take the $s_{11}=(m_2+ m_1)^2$ as an example(the equation (4.21) in \cite{Li:2022cbx})
	\begin{align}
		I_2=\frac{d-2}{2(d-3)m_2(m_1+m_2)}I_{2;\widehat{1}}+\frac{d-2}{2(d-3)m_1(m_1+m_2)}I_{2;\widehat{2}}\ed
	\end{align}
	Substitute the above equation into \eqref{eq:bubble-rank1} and \eqref{requal1}
	\begin{align}
		I_2^{(1)}=\frac{((d-2)m_1+(d-3)m_2)s_{01}}{(d-3)m_2(m_1+m_2)^2}I_{2;\widehat{1}}+\frac{s_{01}}{(d-3)(m_1+m_2)^2}I_{2;\widehat{2}}\co
	\end{align}
	which is consistent with the \cite{Li:2022cbx}.
	 Similarly,
	 \begin{align}
		I_2^{(2)}=\frac{2}{(d-3)(d-1)m_2(m_1+m_2)^3}\left( N_{2;\widehat{1}}^{(1)}I_{2;\widehat{1}} +N_{2;\widehat{2}}^{(1)}I_{2;\widehat{2}}\right)\co
	 \end{align}
	 where
	 \begin{align}N_{2;\widehat{1}}^{(1)}=&(d-2) s_{01}^2 \left[(d-1) m_{1}^2+(d-3) m_{2}^2\right]+(d-3) m_{2} \left[2 (d-1) m_{1} s_{01}^2+m_{2} s_{00} (m_{1}+m_{2})^2\right]\co\nn
	 	N_{2;\widehat{2}}^{(1)}=& m_{1}m_2  \left[(d-3) s_{00} (m_{1}+m_{2})^2+2 s_{01}^2\right]\ed
	 \end{align}
	
	\subsubsection{$I_{\{2,1\}}$ and $I_{\{2,1\}}^{(1)}$}
	
	For clarity, we calculate the $I_{\{2,1\}}$ using \eref{highrecur2}
	\begin{align}
		&\textsf{H}^{1+}_2 I_{\{2,1\}} + \textsf{H}_2 I_{\{1,1\}}  + \textsf{H}_{2;\widehat{1}} I_{\{0,1\}} +\Big[\left(H_{2;\widehat{2}}+\textsf{H}^{1+}_{2;\widehat{22}}\right) I_{\{1,0\}}\nn
		&+ \Big((m_2^2-m_1^2+q_2^2)\textsf{H}^{1+}_{2;\widehat{22}}+\textsf{H}^{1+}_{2;\widehat{2}}\Big) I_{\{2,0\}}-\textsf{H}^{1+}_{2;\widehat{22}} q_2\cdot \d_R I_{\{2,0\}}^{(1)}\Big]=0\ed
	\end{align}
	Based on the results derived in the preceding section,
	\begin{align}
		I_{\{2,0\}}={d-2\over 2 m_1^2} I_{\{1,0\}},~ I_{\{2,0\}}^{(1)}=0\ed
	\end{align}
	We can get the final result of $I_{\{2,1\}}$
	\begin{align}
		&I_{\{2,1\}}\nn
		&=- {\textsf{H}_2 \over \textsf{H}_2^{1+}} I_{\{1,1\}} -{\textsf{H}_{2;\widehat{1}} \over \textsf{H}_2^{1+}} I_{\{0,1\}} -\left( {\textsf{H}_{2;\widehat{2}}+\textsf{H}^{1+}_{2;\widehat{22}} \over \textsf{H}_2^{1+}} + {(m_1^2-m_2^2+s_{11})\textsf{H}^{1+}_{2;\widehat{22}}+\textsf{H}^{1+}_{2;\widehat{2}} \over \textsf{H}_2^{1+}} {d-2\over 2 m_1^2}\right) I_{\{1,0\}} \nn
		&=\frac{(d-3) \left(m_1^2-m_2^2-s_{11}\right)}{4\textsf{H}_2^{1+}} I_{\{1,1\}}+\frac{d-2}{4\textsf{H}_2^{1+}} I_{\{0,1\}}+\frac{(2-d) \left(m_1^2+m_2^2-s_{11}\right)}{8 m_1 ^2 \textsf{H}_2^{1+}} I_{\{1,0\}}\ed  \label{bubhighpole}
	\end{align}
	where  the second equality holds since we have used the coefficients expression
	\begin{align}
		\textsf{H}_2^{1+}&=\frac{1}{4} \left(-2 m_1^2 \left(m_2^2+s_{11}\right)+\left(m_2^2-s_{11}\right){}^2+m_1^4\right),\nn
		\textsf{H}_2&=-\frac{1}{4} (d-3) \left(m_1^2-m_2^2-s_{11}\right), ~  \textsf{H}_{2;\widehat{1}}=\frac{2-d}{4}, \nn
		\textsf{H}^{1+}_{2;\widehat{2}} &=\frac{1}{2} \left(-m_1^2+m_2^2-s_{11}\right), ~ \textsf{H}_{2;\widehat{2}}=\frac{d-3}{4}, ~\textsf{H}^{1+}_{2;\widehat{22}} =\frac{1}{4}\ed
	\end{align}
	Acting $\d_{m_1^2}$ on $I_{\{1,1\}}^{(1)}$ can be derived using the \eref{bubhighrank}-\eref{requal1},
	\begin{align}
		I_{\{2,1\}}^{(1)}={s_{01}\over s_{11}}I_{\{1,1\}}+{s_{01}(m_1^2-m_2^2+s_{11})\over s_{11}}I_{\{2,1\}}-{s_{01}\over s_{11}}{d-2\over 2m_1^2}I_{\{1,0\}} \ed
	\end{align}
	Plugging \eref{bubhighpole} into it,
	\begin{align}
		I_{\{2,1\}}^{(1)}=&\frac{s_{01} \Big[s_{11} \Big((d-4) \left(m_2^2-s_{11}\right)-2 m_1^2\Big)+(d-2) \Big((m_1^2-m_2^2)^2 -m_2^2 s_{11}\Big)\Big]}{4 s_{11}  \textsf{H}_2^{1+}}I_{\{1,1\}}\nn
		&+\frac{(d-2) s_{01} \left(m_1^2-m_2^2+s_{11}\right)}{4 s_{11} \textsf{H}_2^{1+}} I_{\{0,1\}}+\frac{(2-d) s_{01} \left(m_1^2-m_2^2-s_{11}\right)}{4 s_{11} \textsf{H}_2^{1+}} I_{\{1,0\}}\ed
	\end{align}
	
	\subsection{Triangles}
	
	The preceding subsections have provided detailed and specific computations for the tadpole and bubble. Consequently, in what follows, the step-by-step calculations will be omitted for brevity. Unlike tadpoles and bubbles, triangles involve integrals where two propagators have been removed. Using the  \eref{firstresult}, we can get the triangle equation,
	\begin{align}
		\textsf{A}_{3,r} I_n^{(r)} +\textsf{B}_{3,r} I_3^{(r-1)} +\textsf{C}_{3,r}I_3^{(r-2)} +\textsf{B}_{3,r;\widehat{i}} I_{3;\widehat{i}}^{(r-1)}+\textsf{C}_{n,r;\widehat{i}}I_{3;\widehat{i}}^{(r-2)}+\sum_{i<j}\textsf{C}_{3,r;\widehat{ij}}I_{3;\widehat{ij}}^{(r-2)}=0
	\end{align}
	Since only $\mathcal{G}_{00}$, $\mathcal{G}_0$ and $\mathcal{G}_c$ are associated with singularities, we will focus exclusively on presenting their specific results. The other coefficients in the reduction formula do not contribute to singular configurations, and hence will be omitted for brevity.
	\begin{align}
		\mathcal{G}_{00}=&\frac{1}{4}(s_{11} s_{22}-s_{12}^2)\co\\ \label{tg00}
		\mathcal{G}_0=&\frac{1}{2} \left(m_1^2 \left(s_{12}-s_{22}\right)-m_3^2 s_{12}+s_{22} \left(m_2^2-s_{11}+s_{12}\right)\right) s_{01}\nonumber\\
		&+\frac{1}{2} \left(m_1^2 \left(s_{12}-s_{11}\right)+m_3^2 s_{11}-m_2^2 s_{12}+s_{11} s_{12}-s_{11} s_{22}\right) s_{02}\co\\ \label{tg0}
		\mathcal{G}_c=&-m_1^2  s_{12}^2 s_{00}+\frac{1}{4}\left(  s_{22}^2 s_{01}^2+  \left(m_1^2 \left(s_{01}-s_{02}\right)-m_3^2 s_{01}+m_2^2 s_{02}\right){}^2+ s_{11}^2 \left( s_{02}^2-  s_{22} s_{00}\right)\right)\nn
		&-\frac{s_{22}}{4} \left(m_1^4 s_{00}+m_1^2 \left(2 s_{01} \left(s_{01}+s_{02}\right)-2 m_2^2 s_{00}\right)+2 m_3^2 s_{01}^2+m_2^4 s_{00}-2 m_2^2 s_{01} s_{02}\right)\nn
		&+\frac{s_{12}}{2} \left( \left(m_1^2-m_2^2\right)  s_{22} s_{00}+ \left(m_1^4 s_{00}-m_1^2 \left(m_2^2 s_{00}+m_3^2 s_{00}-4 s_{01} s_{02}\right)+m_2^2 m_3^2 s_{00}\right)\right)\nn
		&+s_{11} \left(\frac{1}{2}  s_{22} \left(m_2^2 s_{00}+m_3^2 s_{00}-s_{01} s_{02}\right)+\frac{s_{12}}{2} \left( \left(m_1^2-m_3^2\right)  s_{00}+ s_{22} s_{00}\right)-\frac{1}{4}  s_{22}^2 s_{00}\right)\nn
		&-\frac{s_{11}}{4}  \left(m_1^4 s_{00}+m_1^2 \left(2 s_{02} \left(s_{01}+s_{02}\right)-2 m_3^2 s_{00}\right)+2 m_2^2 s_{02}^2+m_3^4 s_{00}-2 m_3^2 s_{01} s_{02}\right)\ed
	\end{align}
	First, the case of $\mathcal{G}_{00}\neq0$ will be considered. Following a procedure analogous to that applied to the Bubble, the final result for the triangle can be derived.
	\begin{align}
		I_3^{(r)}= C_{3}^{(r)}I_3+\sum_{i=1,2,3}C_{3;\widehat{i}}^{(r)}I_{3;\widehat{i}}+\sum_{1 \leq i<j\leq 3}C_{3;\widehat{ij}}^{(r)}I_{3;\widehat{ij}}\ed
	\end{align}
	The specific coefficients for different values of $r$:
	\begin{itemize}
		\item $r=1$
		\begin{align}\label{eq:triangle-rank1}
			C_{3}^{(1)}=&-\frac{\mathcal{G}_{0}}{2 \mathcal{G}_{00}}\co \nn
			C_{3;\widehat{3}}^{(1)}=&\frac{s_{12} s_{01}-s_{11} s_{02}}{4 \mathcal{G}_{00}}\co \quad C_{3;\widehat{2}}^{(1)}=\frac{s_{12} s_{02}-s_{22} s_{01}}{4 \mathcal{G}_{00}}\co \nn
			C_{3;\widehat{1}}^{(1)}=&\frac{\left(s_{22}-s_{12}\right) s_{01}-\left(s_{12}-s_{11}\right) s_{02}}{4 \mathcal{G}_{00}}\co
		\end{align}
		\item $r=2$
	\end{itemize}
	In this case, the analytic expressions are unwieldy, so a numerical solution will be adopted. To facilitate comparison with prior results from \cite{Li:2022cbx}, the parameters are set as $d=4$, $\{m_1^2,m_2^2,m_3^2\}=\{\frac{1}{2},\frac{1}{3},\frac{1}{5}\},\{s_{12},s_{11},s_{22}\}=
	\{\frac{7}{13},\frac{5}{7},\frac{7t}{5}+\frac{343}{845}\}$, where $t=4\mathcal{G}_{00}$.
	\begin{align}
		C_{3}^{(2)}=&\frac{-1573040 s_{00}-476472423 s_{01}^2+1084364190 s_{02} s_{01}-596126375 s_{02}^2}{2519080200 t}\nonumber\\
		&+\frac{5618 \left(49 s_{01}-65 s_{02}\right)^2}{17738523075 t^2}
		+\mathcal{O}(t^0)\co\\
		C_{3;\widehat{3}}^{(2)}=&\frac{530 s_{00}+24787 s_{01}^2-5200 s_{02} s_{01}-38025 s_{02}^2}{35490 t}-\frac{53 \left(49 s_{01}-65 s_{02}\right)^2}{6997445 t^2}+\mathcal{O}(t^0)\co\\
		C_{3;\widehat{23}}^{(2)}=&\frac{-2401 s_{01}^2+6370 s_{02} s_{01}-4225 s_{02}^2}{3185 t}+\mathcal{O}(t^0)\ed
	\end{align}
	\subsubsection{$ \mathcal{G}_{00}=0$}
	
	There are many viable options to satisfy the condition $\mathcal{G}_{00}=0$; without loss of generality, we impose the constraint $s_{11}=s_{12}^2/s_{22}$. As with the bubbles, in this subsection $I_3^{(r)}$ represents the modified version of the integral. Using \eref{thirdresult}, the final result for $I_3^{(r)}$ is
	\begin{align}
		I_3^{(r)}= \sum_{i=1,2,3}C_{3;\widehat{i}}^{(r)}I_{3;\widehat{i}}+\sum_{1 \leq i<j\leq 3}C_{3;\widehat{ij}}^{(r)}I_{3;\widehat{ij}}\ed
	\end{align}
	The  coefficients for different values of $r$:
	\begin{itemize}
		\item $r=0$
		\begin{align}
			C_{3;\widehat{3}}=&\frac{s_{12}}{m_1^2 \left(s_{12}-s_{22}\right)-m_3^2 s_{12}+m_2^2 s_{22}-s_{12}^2+s_{12} s_{22}}\co\\
			C_{3;\widehat{2}}=&\frac{s_{22}}{m_1^2 \left(s_{12}-s_{22}\right)-m_3^2 s_{12}+m_2^2 s_{22}-s_{12}^2+s_{12} s_{22}}\co\\
			C_{3;\widehat{1}}=&\frac{s_{22}-s_{12}}{m_1^2 \left(s_{12}-s_{22}\right)-m_3^2 s_{12}+m_2^2 s_{22}-s_{12}^2+s_{12} s_{22}}\co
		\end{align}
		\item $r=1$ 
		\begin{align}
			C_{3;\widehat{3}}^{(1)}=-\frac{169 \left(568463 s_{01}-972595 s_{02}\right)}{943824}\co\quad C_{3;\widehat{23}}^{(1)}=-\frac{169}{371} \left(49 s_{01}-65 s_{02}\right),
		\end{align}
	\end{itemize}
	where we use the same parameter sets in this subsection beginning. 
	
	\subsubsection{$\mathcal{G}_{00}=0$ and $\mathcal{G}_0=0$}
	
	Under the imposed constraint $s_{11}=s_{12}^2/s_{22}$, $\mathcal{G}_0$ and $\mathcal{G}_c$  take the form
	\begin{align}
		\mathcal{G}_0=&\frac{\big(m_1^2 \left(s_{12}-s_{22}\right)-m_3^2 s_{12}+m_2^2 s_{22}-s_{12}^2+s_{12} s_{22}\big) \big(s_{12} s_{02}-s_{22} s_{01}\big)}{2 s_{22}}, \\
		\mathcal{G}_c=&{\mathsf{N}_c \over 4 \left(-m_1^2+m_2^2+s_{12}\right){}^2 \left(-m_1^2+m_3^2+s_{12}\right){}^2},
	\end{align} 
	where 
	\begin{align}
		\mathsf{N}_c=&\left(\left(m_1^2-m_3^2-s_{12}\right) s_{01}+\left(m_2^2-m_1^2+s_{12}\right) s_{02}\right){}^2  \left((m_1+m_2) (m_1-m_3)-s_{12}\right)\nn
		& \times\left((m_1-m_2) (m_1+m_3)-s_{12}\right) \left((m_1-m_2) (m_1-m_3)-s_{12}\right)\nn 
		&\times \left((m_1+m_2) (m_1+m_3)-s_{12}\right).
	\end{align}
	To satisfy $\mathcal{G}_0=0$, we set
	\begin{align}
		s_{22}= -\frac{m_3^2 s_{12}-m_1^2 s_{12}+s_{12}^2}{m_1^2-m_2^2-s_{12}}\ed
	\end{align}
	Clearly, utilizing alternative constraints is also valid, as the underlying algorithm remains unchanged. The final result for $I_3^{(r)}$ can be expressed as
	\begin{align}
		I_3^{(r)}= \sum_{i=1,2,3}C_{3;\widehat{i}}^{(r)}I_{3;\widehat{i}}+\sum_{1 \leq i<j\leq 3}C_{3;\widehat{ij}}^{(r)}I_{3;\widehat{ij}}\ed
	\end{align}
	The  coefficients for different values of $r$:
	\begin{itemize}
		\item $r=0$ 
		\begin{align}
			C_{3;\widehat{3}}=&{(3-d) \left(m_1^2-m_2^2-s_{12}\right) \left(m_1^2-m_3^2-s_{12}\right) \left(\left(m_3^2-m_1^2\right) (m_2^2-m_1^2)-s_{12}^2\right) s_{01} \over \mathsf{D}_{3;\widehat{3}}}, \nn
			C_{3;\widehat{23}}=&{(d-2)\left(m_1^2-m_2^2-s_{12}\right) \left(m_1^2-m_3^2-s_{12}\right)  \over \mathsf{D}_{3;\widehat{23}}},
		\end{align}
		where
		\begin{align}
			\mathsf{D}_{3;\widehat{3}}=&\left(\left(m_1^2-m_3^2-s_{12}\right) s_{01}+\left(m_2^2-m_1^2+s_{12}\right) s_{02}\right) \left((m_1+m_2) (m_1-m_3)-s_{12}\right)\nn
			& \times\left((m_1-m_2) (m_1+m_3)-s_{12}\right) \left((m_1-m_2) (m_1-m_3)-s_{12}\right)\nn 
			&\times \left((m_1+m_2) (m_1+m_3)-s_{12}\right),\nn
			\mathsf{D}_{3;\widehat{23}}=& \left((m_1+m_2) (m_1-m_3)-s_{12}\right)\left((m_1-m_2) (m_1+m_3)-s_{12}\right)\nn &\times\left((m_1-m_2) (m_1-m_3)-s_{12}\right) \left((m_1+m_2) (m_1+m_3)-s_{12}\right).
		\end{align}
	\end{itemize}
	Specific results for rank $r\geq1$ are omitted here in the interest of brevity, but the computational procedure remains unchanged.
	As evident from (\ref{tg00}) and (\ref{tg0}), setting $q_3=0$ results in both $\mathcal{G}_{00}$ and $\mathcal{G}_0$ vanishing. In this case, the  coefficients for different values of $r$:
	\begin{itemize}
		\item $r=0$
	\begin{align}
			C_{3;\widehat{1}}&= -\frac{1}{m_1^2-m_3^2}, ~C_{3;\widehat{3}}=\frac{1}{m_1^2-m_3^2}, ~C_{3;\widehat{2}}=0, \nn
			C_{3;\widehat{23}}&= 0, ~C_{3;\widehat{13}}=0, ~ C_{3;\widehat{12}}=0,
		\end{align}
		\item $r=1$
		\begin{align}
			&C_{3;\widehat{3}}^{(1)}=\frac{\left(m_1^2-m_2^2+s_{11}\right) \left(2 m_1^2-m_2^2-m_3^2+s_{11}\right) s_{01}}{\left(m_1^2-m_3^2\right){}^2 s_{11}}, ~ C_{3;\widehat{2}}^{(1)}=0,~ \nn  &C_{3;\widehat{1}}^{(1)}=-\frac{\left(m_1^2+m_2^2-2 m_3^2-s_{11}\right) \left(-m_2^2+m_3^2+s_{11}\right) s_{01}}{\left(m_1^2-m_3^2\right){}^2 s_{11}}, \nn
			&C_{3;\widehat{23}}^{(1)}=\frac{\left(-2 m_1^2+m_2^2+m_3^2-s_{11}\right) s_{01}}{\left(m_1^2-m_3^2\right){}^2 s_{11}}, ~ C_{3;\widehat{13}}^{(1)}=\frac{\left(m_1^2-2 m_2^2+m_3^2+2 s_{11}\right) s_{01}}{\left(m_1^2-m_3^2\right){}^2 s_{11}}, ~ \nn
			&C_{3;\widehat{12}}^{(1)}=\frac{\left(m_1^2+m_2^2-2 m_3^2-s_{11}\right) s_{01}}{\left(m_1^2-m_3^2\right){}^2 s_{11}}\ed
		\end{align}
	\end{itemize}
	The results of integrals with tensor structure in the examples examined herein match those derived in \cite{Li:2022cbx}. 
	
	\subsubsection{${{\rm Det} ~Q=0}$}
	
	 Analogous to the bubble, using the scalar result in \cite{Li:2022cbx}. The parameters are set as  $\{m_1^2,m_2^2,m_3^2\}=\{\frac{1}{2},\frac{1}{3},\frac{5}{338}\},\{s_{11},s_{12},s_{22}\}=
	 \{\frac{5}{7},\frac{7}{13},\frac{3552}{5915}\}$, 
	 \begin{align}
	 	I_3=&\frac{21}{1151(d-4)}\Big[4225(d-2)I_{3;\widehat{12}}-780(d-2)I_{3;\widehat{13}}+455(d-2)I_{3;\widehat{23}}\nn
	 	&+402(d-3)I_{3;\widehat{1}}-592(d-3)I_{3;\widehat{2}}+130(d-3)I_{3;\widehat{3}}\Big]\ed
	 \end{align}
	Substitute the above equation into \eqref{eq:triangle-rank1}, 
	\begin{align}
		I_{3}^{(1)}&=\frac{7}{11510 (d-4)} \Bigg[-4225 (d-2) \left(12 s_{01}-65 s_{02}\right)I_{3;\what{12}} +780 (d-2) \left(12 s_{01}-65
		s_{02}\right) I_{3;\what{13}}\nn
		&\newline -455 (d-2) \left(12 s_{01}-65 s_{02}\right) I_{3;\what{23}}+130
		\left((37 d-160) s_{01}+65 s_{02}\right)I_{3;\what{3}}\nn
		&\newline +2  \left(65 (692-247 d) s_{02}+3552
		s_{01}\right)I_{3;\what{2}}-2  \left((2045 d-5768) s_{01}+65 (667-217 d) s_{02}\right)I_{3;\what{1}}\Bigg]\ed
	\end{align}
	\subsubsection{$I_{\{2,1,1\}}$}
	
	The integral $I_{\{2,1,1\}}$ is evaluated using \eref{highrecur2}
	\begin{align}
		&\textsf{H}^{1+}_3 I_{\{2,1,1\}} + \textsf{H}_3 I_{\{1,1,1\}}  + \textsf{H}_{3;\widehat{1}} I_{\{0,1,1\}} +\sum_{j=2,3 }\Big[\left(\textsf{H}_{3;\widehat{j}}+\textsf{H}^{1+}_{3;\widehat{jj}}\right) I_{\{a_j=0\}}\nn
		&+ \Big((m_1^2-m_j^2+q_j^2)\textsf{H}^{1+}_{3;\widehat{jj}}+\textsf{H}^{1+}_{3;\widehat{j}}\Big) I_{\{a_1=2,a_j=0\}}-\textsf{H}^{1+}_{3;\widehat{jj}} q_j\cdot \d_R I_{\{a_1=2,a_j=0\}}^{(1)}\Big]=0\ed 
	\end{align}
	Noting that the specific integrals $I_{\{a_1=2,a_j=0\}}$ and $I_{\{a_1=2,a_j=0\}}^{(1)}$ follow from  corresponding bubbles section. 
	\allowdisplaybreaks
	\begin{align}
		I_{\{2,1,1\}} = C_{3\to3}^{1+} I_{\{1,1,1\}} + \sum_{i=1,2,3}C_{3\to3;\widehat{i}}^{1+} I_{\{a_i=0\}} + \sum_{i\neq j}C_{3\to3;\widehat{ij}}^{1+} I_{\{a_i=0,a_j=0\}}\co
	\end{align}
	where the coefficients are
	\begin{align}
		C_{3\to3}^{1+}&= {(d-4) \left[(m_1^2-s_{12}) \left(s_{11}-2 s_{12}+s_{22}\right)+m_3^2 \left(s_{12}-s_{11}\right)+m_2^2 \left(s_{12}-s_{22}\right)\right]\over 4 B_3^{1+}} , \nn
		C_{3\to3;\widehat{1}}^{1+}&= {(d-3) \left(s_{11}-2 s_{12}+s_{22}\right) \over 4 B_3^{1+}} ,\nn
		C_{3\to3;\widehat{2}}^{1+}&= {\mathsf{N}_{3\to3;\widehat{2}}\over 4 (m_1^4-2 m_1^2 \left(m_3^2+s_{22}\right)+\left(m_3^2-s_{22}\right){}^2) B_3^{1+}} ,\nn
		C_{3\to3;\widehat{3}}^{1+}&= {\mathsf{N}_{3\to3;\widehat{3}}\over 4 (m_1^4-2 m_1^2 \left(m_2^2+s_{11}\right)+\left(m_2^2-s_{11}\right){}^2) B_3^{1+}} ,\nn
		C_{3\to3;\widehat{12}}^{1+}&= {(d-2) \left(m_1^2 \left(s_{22}-s_{12}\right)+m_3^2 s_{12}-s_{22} \left(m_2^2-s_{11}+s_{12}\right)\right)\over {4 (m_1^4-2 m_1^2 \left(m_3^2+s_{22}\right)+\left(m_3^2-s_{22}\right){}^2) B_3^{1+}}} ,  \nn
		C_{3\to3;\widehat{13}}^{1+}&= {(d-2) \left(m_1^2 \left(s_{11}-s_{12}\right)+m_2^2 s_{12}-s_{11} (m_3^2- s_{22}+s_{12})\right) \over4 (m_1^4-2 m_1^2 \left(m_2^2+s_{11}\right)+\left(m_2^2-s_{11}\right){}^2) B_3^{1+}} ,  \nn
		C_{3\to3;\widehat{23}}^{1+}&={\mathsf{N}_{3\to3;\widehat{23}}\over \mathsf{D}_{3\to3;\widehat{23}} },
	\end{align}
	where
	\begin{align}
		B_3^{1+}=& \frac{1}{2} \left(s_{22}-s_{12}\right) \left(-m_3^2 s_{11}-m_1^2 m_2^2\right)+\frac{1}{4} \left(s_{11}-2 s_{12}+s_{22}\right) \left(-2 m_1^2 s_{12}+m_1^4+s_{11} s_{22}\right)\nn
		&+\frac{1}{4} \left(m_3^4 s_{11}-2 m_2^2 m_3^2 s_{12}+2 m_1^2 m_3^2 \left(s_{12}-s_{11}\right)+m_2^2 s_{22} \left(m_2^2-2 s_{11}+2 s_{12}\right)\right),\nn
		\mathsf{N}_{3\to3;\widehat{2}}=&(3-d) s_{22}\Big[\left(m_3^2-m_1^2\right) \left(-m_2^2+m_3^2+s_{11}\right)+2 s_{12} \left(m_1^2+m_3^2-s_{22}\right)\nn
		&+s_{22} \left(-m_1^2-m_2^2-2 m_3^2+s_{11}+s_{22}\right)\Big],	\nn
		\mathsf{N}_{3\to3;\widehat{3}}=&(3-d) s_{11}\Big[\left(m_1^2-m_2^2\right) \left(-m_2^2+m_3^2-s_{22}\right)+2 s_{12} \left(m_1^2+m_2^2-s_{11}\right)\nn
		&+s_{11} \left(-m_1^2-2 m_2^2-m_3^2+s_{11}+s_{22}\right)\Big],	\nn
		\mathbf{N}_{3\to3;\widehat{23}}=& (d-2)\Big\{\left[(m_1^2-m_2^2){}^3+s_{11}^3\right](s_{22}-m_1^2-m_3^2)s_{22}\nn
		&+2 s_{12} \left(m_1^2-m_2^2\right) \left[m_1^6+m_2^2 \left(m_3^2-s_{22}\right){}^2-m_1^2(m_1^2+m_2^2) \left(m_3^2+s_{22}\right)\right]\nn
		&+s_{11}^2\big[2 m_3^2 \left(m_1^2-m_3^2\right) s_{12}+3 \left(m_2^2+m_3^2\right) s_{22} \left(m_3^2-s_{22}\right)\nn
		&+s_{22} \left(m_1^2 \left(3 m_2^2-m_3^2+2 s_{12}-2 s_{22}\right)+4 m_3^2 s_{12}+s_{22} \left(s_{22}-2 s_{12}\right)\right)+\left(m_1^2-m_3^2\right){}^3\big]\nn
		&+s_{11}\Big[\left(3 m_2^2 \left(m_2^2+m_3^2\right)-m_1^2 \left(m_2^2-3 m_3^2\right)\right) s_{22}^2-\left(m_1^2+m_2^2\right) (\left(m_1^2-m_3^2\right){}^3+s_{22}^3)\nn
		&+\left(2 m_1^4-3 m_2^4-3 m_3^4+4 m_2^2 m_3^2\right)s_{22} m_1^2+3s_{22} \left(m_2^2+m_3^2\right) \left(m_1^4-m_2^2 m_3^2\right)\nn
		&+s_{12} \left(2 m_1^2 \left(\left(m_3^2-s_{22}\right){}^2-2 m_2^2 \left(m_3^2+s_{22}\right)\right)+4 m_2^2 \left(m_3^2-s_{22}\right){}^2-2 m_1^6\right) \Big]\Big\}, \nn 
		\mathsf{D}_{3\to3;\widehat{23}}=& 8 m_1^2 \left(\left(m_1-m_2\right){}^2-s_{11}\right) \left(\left(m_1+m_2\right){}^2-s_{11}\right) \nn
		&\times \left(\left(m_1-m_3\right){}^2-s_{22}\right) \left(\left(m_1+m_3\right){}^2-s_{22}\right) B_3^{1+}.
	\end{align}
	The integrals with higher poles presented in the tadpoles, bubbles, and triangles sections are consistent with \cite{Feng:2022uqp}.
	
	\section{Summary and Outlook}
	\label{summary}
	This work has demonstrated a unifying framework that synergizes the Baikov representation and IBP relations to uniformly reduce one-loop integrals with arbitrary tensor structures and high poles. Although our recursion formula includes a term with $\d_R$, this poses little difficulty given the simplicity it provides in avoiding tedious algebraic manipulations. Most importantly, one can easily and consistently treat various degenerate cases appearing in our method. The degeneracy of $\det Q=0$ in \cite{Li:2022cbx} may not be immediately apparent using our method. However, it is worth noting that our degenerate origin, represented by $\mathcal{G}_{00}$ and $\mathcal{G}_0$, does not vanish, our recursion relation remains valid. Although their tensor reduction cannot be effectively handled by our algorithm, we can utilize their scalar reduction result as an input to obtain the tensor reduction.
	
	To restore the general tensor structure in the tensor reduction of $L$-loop integrals, it is necessary to introduce $L$ auxiliary vectors. In contrast to the one-loop case, the inclusion of ISP's such as $\ell_i\cdot R_j$ and $\ell_i\cdot p_j$ becomes necessary.  We call the ISP $\ell_i\cdot R_i$ as R-ISP as they emerge in the momentum representation. In general, we can derive $L$ recursion relations by considering differentiation with respect to the $L$ R-ISP's.\footnote{In the subsequent discussion, unless explicitly indicated, the term ``ISP" refers to the ISP's excluding R-ISP's.}. However, these relations alone are insufficient due to the presence of ISP's in $\partial_{z_{0i}}\mathcal{G}(z)$ and $\mathcal{G}(z)$, where $z_{0i}=\ell_i \cdot R_i$. Unfortunately, there is no established method for handling these terms effectively. One approach is to translate these terms into differential operators acting on the auxiliary vectors. Consequently, unlike the one-loop case, we obtain differential equations instead of pure recursion relations. To solve the tensor reduction problem, one can consider expanding the reduction coefficient based on its tensor structure, which leads to $L$ recursion relations. Next, we introduce a linear combination of propagators in the numerator by applying differential operators. This gives rise to $N$ recursion relations, where $N$ is the number of propagators. Assuming there are $E$ independent external momenta, if $L+N<{L(L+1)\over 2}+EL$, we can consider applying $\partial_{z_{\text{ISP}}}$ to generate additional required recursion relations. In this way, we can ultimately solve the tensor reduction problem. However, it should be noted that redundancy may arise when the number of ISP's is large compared to the number of extra recursion equations needed.
	
	To illustrate this process, let's consider the sunset diagram as an example. We introduce two auxiliary vectors, $R_1$ and $R_2$, which results in more ISP's involved in the Baikov representation, namely $\ell_1\cdot p$, $\ell_2\cdot p$, $\ell_1\cdot R_2$, and $\ell_2\cdot R_1$. There are 5 tensor structures involving $R_1$ and $R_2$, i.e., $R_1^2,R_2^2,R_1\cdot R_2,R_1\cdot p,R_2\cdot p$. The sunset has exactly 2 R-ISP's and 3 propagators.
	Fortunately, we are fortunate enough to solve the reduction problem by utilizing a set of $L+N=5$ recursion relations which are derived through the loop-by-loop reduction and constructing propagators \cite{Feng:2022iuc}. Here we can simply make use of the IBP relations generated by $\partial_{z_{0i}}$:
	\begin{align}
		\int \partial_{z_{0i}}\left[{\mathcal{G}(z)^\gamma z_{0i}^{r_i}\over \prod_jz_j}\right]=0\ed
	\end{align}
	During the reduction process, one may come across terms in the numerator that involve $z_{\text{ISP}}$ and $z_j$. The presence of terms involving $z_j$ allows for a reduction of the integral to a known sector with a lower topology. On the other hand, handling the terms containing $z_{\text{ISP}}$ is relatively straightforward, as they can be readily translated into differential operators acting on $R_{1}$ and $R_{2}$. So finally we obtain two partial differentials of the standard integral $I_{1,1,1}^{(r_1,r_2)}$. The remaining three recursion relations are obtained by constructing propagators in the numerator through the application of three differential operators:
	\begin{align}
		\partial_{R_1}\cdot \partial_{R_1}, \partial_{R_2}\cdot \partial_{R_2},\partial_{R_1}\cdot \partial_{R_2}\ed
	\end{align}
	It is easy to find
	\begin{align}
		\begin{aligned}
			\partial_{R_1}\cdot \partial_{R_1}I_{1,1,1}^{(r_1,r_2)}&=4r_1(r_1-1)\left[m_1^2I_{1,1,1}^{(r_1-2,r_2)}+I_{0,1,1}^{(r_1-2,r_2)}\right] \co\\
			\partial_{R_2}\cdot \partial_{R_2}I_{1,1,1}^{(r_1,r_2)}&=4r_2(r_2-1)\left[m_2^2I_{1,1,1}^{(r_1,r_2-2)}+I_{1,0,1}^{(r_1,r_2-2)}\right] \co\\
			\partial_{R_1}\cdot \partial_{R_2}I_{1,1,1}^{(r_1,r_2)}
			&=2r_1r_2\left[I^{(r_1-1,r_2-1)}_{1,1,1;\what{3}-\what{1}-\what{2}}-(p^2+m_1^2+m_2^2-m_3^2)I^{(r_1-1,r_2-1)}_{1,1,1}\right ]\\
			&\hspace{15pt}+2r_2p\cdot \partial_{R_1}I^{(r_1,r_2-1)}_{1,1,1}+2r_1{p\cdot \partial_{R_2}}I^{(r_1-1,r_2)}_{1,1,1}\ed 
		\end{aligned}
	\end{align}
	To convert the aforementioned differential equations into recursion relations, one can expand the reduction coefficient $C^{(r_1,r_2)}_{a}$ in $I^{(r_1,r_2)}_{1,1,1}=\sum_{a=1}^7 C^{(r_1,r_2)}_{a} I_{a}$
	based on the tensor structure of $R_1$ and $R_2$. 
	\begin{align}
		C^{(r_1,r_2)}_{a}=\sum_{\{v\}}c^{a;(r_1,r_2)}_{\nu_1\nu_2\nu_3\nu_4\nu_5}(R_1^2)^{\nu_1}(R_2^2)^{\nu_2}(R_1\cdot R_2)^{\nu_3}(R_1\cdot p)^{\nu_4}(R_2\cdot p)^{\nu_5}\co
	\end{align}
	with
	\begin{align}
		2\nu_1+\nu_3+\nu_4=r_1\, , \quad  2\nu_2+\nu_3+\nu_5=r_2\ed
	\end{align}
	The five recursion relations of the \textit{expansion coefficients} $c^{a;(r_1,r_2)}_{\nu_1\nu_2\nu_3\nu_4\nu_5}$, as explicitly discussed in \cite{Feng:2022iuc}, provide a complete solution for the tensor reduction of integrals with the sunset topology. For higher-loop tensor integrals with $L+N<{L(L+1)\over 2}+EL$, one can initially generate $L+N$ recursion relations using a similar approach. Subsequently, the remaining recursion relations needed for these cases can be obtained from IBPs generated by taking derivatives with respect to other ISP parameters. As can be anticipated, higher loops will inevitably result in high-degree polynomials. As previously discussed, it is necessary to convert all ISP's in the numerator (excluding the R-ISP's) into differential operators. This transformation can lead to a series of complex partial differential equations with high-order derivatives. In principle, these intricate calculations can be delegated to a computer. However, we must actually resolve the linear equations for the expansion coefficients. Indeed, the method encounters challenges as the total rank $r_1+r_2$ (in the case of two loops) and the number of external momenta increase. This leads to a rapid proliferation in the number of linear equations involved in the reduction process. Improvement for this method for higher loops is left for future research and exploration.
	
	\section*{Acknowledgements}
	
	We are thankful to Bo Feng for useful comments and Tingfei Li for helpful discussions.
	
	\appendix
	\section{Pentagon}\label{sec:appendix}
	
	\begin{table}[ht]
		\centering
		\begin{tabular}{|c|c|c|c|}
			\hline
			& $r$=1 & $r$=2   &$r$=3\\
			\hline
			Our method &  6s&  16s&  39s\\
			\hline
			FIRE6 & 60s& 180s&557s\\\hline
		\end{tabular}
		\caption{The consuming time for different ranks of our method \textit{vs}. FIRE6. In order to expedite the reduction process, we have made the decision to set $m_1=m_2=m_3$ and $m_4=m_5$ during the reduction process in FIRE6. }
		\label{tab:cost_time}
	\end{table}
	In this appendix, we will provide an additional illustrative example of a pentagon. For simplicity,  we will give the numerical result. Let us begin by setting up the numerical framework: $s_{11}= \frac{1}{13},s_{12}= \frac{1}{17},s_{13}= \frac{1}{19},s_{14}= \frac{1}{23}, s_{22}= \frac{2}{29},s_{23}= \frac{2}{31},s_{24}= \frac{2}{37},s_{33}= \frac{3}{41},s_{34}= \frac{3}{43}, s_{44}= \frac{3433409242718675-8300603361746045880868 t}{48244934730591561}$, with $t=\mathcal{G}_{00}$. The definition of $s_{ij}$ remains consistent with the previous formulation. 
	\begin{itemize}
		\item $r=1$
	\end{itemize}
		\begin{align}
			C^{(1)}_{5}=&-\frac{\mathsf{N}^{(1)}_{5}(48324393052 s_{01} - 863984088446 s_{02} + 1984130693427 s_{03} - 
			1318419772377 s_{04})}{10943679594784982799533787019183236\, t}\nn
			&+\mathcal{O}(t^0)\co\\
			C^{(1)}_{5;\widehat{5}}&=\frac{-48324393052 s_{01} + 863984088446 s_{02} - 1984130693427 s_{03} + 
			1318419772377 s_{04}}{226835825478808676\, t}\nn
			&+\mathcal{O}(t^0)\co
		\end{align}
		where
		\begin{align}
			\mathsf{N}^{(1)}_{5}=&165198902691154 + 5487075499569992 m_1^2 + 1768334514951836 m_2^2 \nn
			&- 31615769748504478 m_3^2 + 72605294464574211 m_4^2 - 48244934730591561 m_5^2\ed
		\end{align}
		\begin{itemize}
			\item $r=2$
		\end{itemize}
		 In this case, we additionally assume that all $m_i$ are equal, denoted as $m_i = m$, and we take $d=6$ to avoid excessive complexity. 
		\begin{align}
			C^{(2)}_{5}=&\frac{6822669362590342075877462929~ \mathsf{N_5^{(2)}}}{19960687178885534221068998904386539063038990526693942059576757238616\, t^2}\nn
			&+\mathcal{O}(t^{-1})\co\\
			C^{(2)}_{5;\widehat{5}}&=-\frac{82599451345577~ \mathsf{N_{5}^{(2)}}}{827472864886215336096756770932755381383603943518512\, t^2}+\mathcal{O}(t^{-1})\co\\
			C^{(2)}_{5;\widehat{45}}&=\frac{67~ \mathsf{N_{5}^{(2)}}}{380016026464966029473624813316\, t}+\mathcal{O}(t^0)\ed
		\end{align}
		where
		\begin{align}
			\mathsf{N_5^{(2)}}=&  (48324393052 s_{01} - 
			863984088446 s_{02} + 1984130693427 s_{03} - 1318419772377 s_{04})^2\ed
		\end{align}
	The time required to obtain the results for the tensor rank of a pentagon is presented in Table \ref{tab:cost_time}. The parameter setting remains the same as for the case of $r=1$, without any constraints on the values of $m_i$ and $d$. Our method is a simple \textsc{Mathetica} program, while the FIRE6 algorithm is executed in parallel using a total of 10 computing threads. It is evident that our method offers significantly faster computational efficiency compared to directly solving the IBP relations.

	We can also give the symbolic result for the pentagon using our algorithm. Setting all of the mass equal zero. The time for $r=1$ and $r=2$ are about 6s and 15s respectively.
	\begin{align}
		C_{5}^{(1)}=\frac{\mathsf{N}^{(1)}_{5}}{\mathsf{D}^{(1)}_{5}},\quad C_{5;\widehat{\widehat{5}}}^{(1)}=\frac{\mathsf{N}^{(1)}_{5;\widehat{5}}}{\mathsf{D}^{(1)}_{5}}\co
	\end{align}
	where
	\begin{align}
		\mathsf{N}^{(1)}_{5}=&s_{04} \Big[s_{13} \big(s_{11} (s_{23} s_{24} - s_{22} s_{34})+s_{12} s_{22} s_{34}-2 s_{12} s_{23} s_{44}+s_{12} s_{24} s_{33}+s_{14} s_{22} (s_{23}-s_{33})\big)\nn
		&+s_{11} (s_{12} (s_{23} s_{34}-s_{24} s_{33})+s_{14}( s_{22} s_{33}-s_{23}^2)+s_{22} (s_{24} s_{33}-s_{23} s_{34}+ s_{33} s_{34}- s_{33} s_{44})\nn
		&+s_{23}^2 s_{44}-s_{23} s_{24} s_{33})+s_{12} s_{33} (-s_{12} s_{34}+s_{12} s_{44}-s_{14} s_{22}+s_{14} s_{23})+s_{13}^2 s_{22} (s_{44}-s_{24})\Big]\nn
		&+s_{02} \Big[s_{14} (s_{11} (s_{23} s_{34}-s_{24} s_{33})+s_{12} s_{33} (s_{34}-s_{44})+s_{13} (-2 s_{22} s_{34}+s_{23} s_{44}+s_{24} s_{33}))\nn
		&+s_{13} s_{44} (s_{12} (s_{34}-s_{33})+s_{13} (s_{22}-s_{24}))+s_{11} s_{34} (s_{34} (s_{22}-s_{12})+s_{24} (s_{13}-s_{33}))\nn
		&+s_{11} s_{44} (s_{33} (s_{12}-s_{22}+s_{23}+s_{24})-s_{23}(s_{13} + s_{34}))+s_{14}^2 s_{33} (s_{22}-s_{23})\Big]\nn
		&+s_{03} \Big[s_{14} (s_{11} (s_{23} s_{24}-s_{22} s_{34})+s_{12} (s_{22} s_{34}+ s_{23} s_{44}-2 s_{24} s_{33})+s_{13} s_{22} (s_{24}-s_{44}))\nn
		&+s_{11} s_{44} (-s_{12} s_{23} +s_{22} (s_{13}+s_{23}-s_{33}+s_{34})-s_{23} s_{24})+s_{14}^2 s_{22} (s_{33}-s_{23})\nn
		&+s_{11} s_{24} (s_{34} (s_{12}-s_{22})-s_{13} s_{24}+s_{24} s_{33})+s_{12} s_{44} (s_{12} (s_{33}-s_{34})+s_{13} (s_{24}-s_{22}))\Big]\nn
		&+s_{01} \Big[s_{14} (s_{22} s_{23} s_{34}-s_{22} s_{33} (s_{24}+s_{34}-s_{44})+s_{23} (s_{24} s_{33}-s_{23} s_{44}))\nn
		&+s_{44} (s_{12} s_{33} (s_{22}-s_{23}-s_{24})+s_{12} s_{23} s_{34}-s_{13} s_{22} (s_{23}-s_{33}+s_{34})+s_{13} s_{23} s_{24})\nn
		&+s_{11} \left(s_{22}(s_{34}^2-s_{33} s_{44})+s_{23}(s_{23}s_{44}-2 s_{24} s_{34})+s_{24}^2 s_{33}\right)\nn
		&-(s_{13} s_{24}-s_{12} s_{34}) (s_{24} s_{33}-s_{22} s_{34})\Big]\co\nn
		\mathsf{N}^{(1)}_{5;\widehat{5}}=&-s_{04} \Big[s_{11} \left(s_{23}^2-s_{22} s_{33}\right)+s_{12}^2 s_{33}-2 s_{12} s_{13} s_{23}+s_{13}^2 s_{22}\Big]+s_{03} \Big[s_{11} s_{23} s_{24}-s_{11} s_{22} s_{34}\nn
		&+s_{12}^2 s_{34}-s_{12} s_{13} s_{24}-s_{12} s_{14} s_{23}+s_{13} s_{14} s_{22}\Big]+s_{02} \Big[s_{11} s_{23} s_{34}-s_{11} s_{24} s_{33}+s_{12} s_{14} s_{33}\nn
		&-s_{13} (s_{12} s_{34}+s_{14} s_{23})+s_{13}^2 s_{24}\Big]+s_{01}\Big[s_{12} s_{24} s_{33}-s_{12} s_{23} s_{34}+s_{13} s_{22} s_{34}-s_{13} s_{23} s_{24}\nn
		&+s_{14} \left(s_{23}^2-s_{22} s_{33}\right)\Big]\co \nn
		\mathsf{D}^{(1)}_{5}=&s_{44} \left(-s_{11} s_{22} s_{33}+s_{11} s_{23}^2+s_{12}^2 s_{33}-2 s_{12} s_{13} s_{23}+s_{13}^2 s_{22}\right)+s_{11} s_{22} s_{34}^2-2 s_{11} s_{23} s_{24} s_{34}\nn
		&+s_{11} s_{24}^2 s_{33}-s_{12}^2 s_{34}^2+2 s_{14} (s_{12} (s_{23} s_{34}-s_{24} s_{33})+s_{13} (s_{23} s_{24}-s_{22} s_{34}))\nn
		&+2 s_{12} s_{13} s_{24} s_{34}-s_{13}^2 s_{24}^2+s_{14}^2 \Big(s_{22} s_{33}-s_{23}^2\Big)\ed
	\end{align}
	The result for rank $r=2$ is given in the ancillary file \textbf{pentagon.nb}.
	
	\bibliographystyle{JHEP}
	\bibliography{reference}
\end{document}